\documentclass[letterpaper, 10 pt, conference]{ieeeconf}  

\IEEEoverridecommandlockouts                              
\usepackage[utf8]{inputenc}
\usepackage[T1]{fontenc}
\usepackage{graphicx}

\usepackage{amsmath}
\usepackage{amssymb}
\usepackage{latexsym}
\usepackage{url}

\title{\LARGE \bf
Editing Mesh Sequences with Varying Connectivity
}

\author{Filip Hácha$^{1}$, Jan Dvořák$^{1}$, Zuzana Káčereková$^{1}$, Libor Váša$^{1}$
\thanks{$^{1}$Department of computer science and engineering, Faculty of applied sciences, University of West Bohemia, Univerzitní 8, 301 00 Plzeň, Czech Republic}%
}

\usepackage{hyperref}
\begin{document}

\maketitle
\thispagestyle{empty}
\pagestyle{empty}

\begin{abstract}
Time-varying connectivity of triangle mesh sequences leads to substantial difficulties in their processing. Unlike editing sequences with constant connectivity, editing sequences with varying connectivity requires addressing the problem of temporal correspondence between the frames of the sequence. We present a method for time-consistent editing of triangle mesh sequences with varying connectivity using sparse temporal correspondence, which can be obtained using existing methods.
Our method includes a deformation model based on the usage of the sparse temporal correspondence, which is suitable for the temporal propagation of user-specified deformations of the edited surface with respect to the shape and true topology of the surface while preserving the individual connectivity of each frame. Since there is no other method capable of comparable types of editing on time-varying meshes, we compare our method and the proposed deformation model with a baseline approach and demonstrate the benefits of our framework. 
\end{abstract}
\section{Introduction}
\emph{Time-varying meshes (TVMs)} are an increasingly popular format for representing time-varying 3D surface data. A~TVM consists of a~sequence of frames, each frame being an independent triangle mesh with its own number of vertices/triangles and its own connectivity. Unlike in \emph{dynamic meshes}, which have the same number of vertices and a~constant connectivity for the entire sequence, making it possible to interpret the positions of vertices as trajectories, in TVMs, the temporal correspondence is not known. This makes the processing tasks, such as compression or texturing, much more difficult with TVMs than with dynamic meshes.


In this work, we address a particular processing task: editing. The objective is to provide the user with the means to apply minor changes to the shape (or shape sequence in our case) using natural inputs, such as dragging a~certain part of the surface or marking an another part as static, limiting the influence of the manipulation. With mesh sequences, the user is expected to manipulate a single frame, and the effect should automatically propagate to nearby frames or to the whole sequence, taking into account the character of the movement in order to provide a natural result. With dynamic meshes, this task is much easier, because identifying in each frame which vertices correspond to those modified or marked by the user is trivial. Moreover, the naturally available correspondence information makes the analysis of the movement also much easier. None of this is directly possible with TVMs.

\begin{figure*}[ht]
  \centering
  \includegraphics[width=\textwidth]{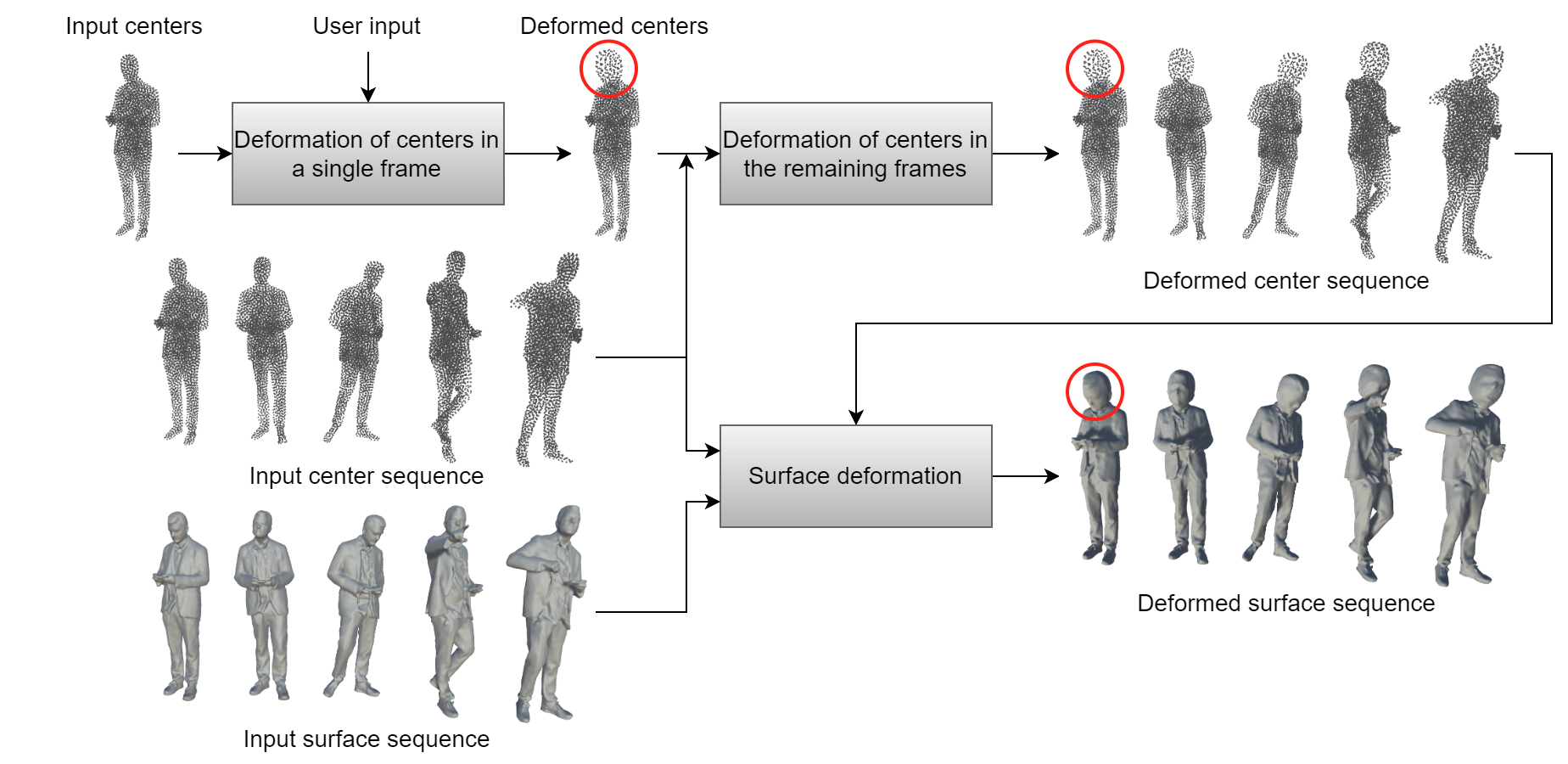}
  \caption{Outline of the proposed editing framework. The edited data shows the volume expansion in the head region and the subsequent time-consistent propagation of the editing to the entire sequence.}
  \label{fig:hl}
\end{figure*}

A potential solution to the problem is finding the surface correspondence for TVMs. However, this turns out to be a~difficult problem, for which there is currently no satisfactory solution. One of the core problems is that due to self-contact, the surface correspondence is not expected to be bijective. Parts of the surface may be hidden in certain frames, and this in turn may lead to frames of different topology, even when the true topology of the captured object remains the same throughout the sequence. This makes optimizing the correspondence very difficult, because it requires distinguishing between a~poor match and a truly non-existent match. 

An attempt to circumvent this problem was made in a~recent series of works \cite{Dvorak2021, Dvorak2022, Dvorak2023}. The authors propose establishing a~correspondence between \emph{elements of volume} enclosed by the surface rather than between surface points, observing that such a~correspondence is bijective for a much wider class of possible input data, making the optimization feasible. Their result, however, is only a discrete (sparse) set of volume elements for which the temporal correspondence is established. It is therefore not straightforward how such information could be used to propagate a~surface modification from one frame to the others. 

In this paper, we propose a framework that allows using the discrete volume correspondence in order to edit a~TVM sequence. First, the user edits a~subset of discrete volume elements (effectors) in a~single frame. Next, three main steps follow: a~consistent deformation of the remaining volume elements in the same frame, the propagation of the deformation to volume elements in the other frames of the sequence, and the deformation of the mesh in the other frames (see Fig. \ref{fig:hl}). The proposed framework is applicable to a~range of editing operations, and it incorporates information from the whole input sequence, making the result more plausible. In particular, by analyzing the correspondence data, it is possible to distinguish between parts of the shape that are truly connected and those that are merely in a~temporary contact. This distinction is in turn essential for providing a~natural result.

In summary, our contributions are:
\begin{itemize}
    \item Building a~complete pipeline for time-consistent editing of time-varying meshes.
    \item Adapting the editing pipeline to use dual quaternions as a~representation of rigid transformations and showing their advantages over the description by translation vectors.
    \item Proposing a~deformation model based on using the affine neighborhood of the edited point, which respects the actual topology of the edited surface instead of the near neighborhood of the point.
\end{itemize}


\section{Related work}
In the past, the main focus in mesh editing was on a~single triangle mesh. The objective in this setting is to alter the mesh geometry, given specified constraints, without introducing significant distortion. Most of the mesh editing methods can be classified into two groups (see Chapter~9 of~\cite{pmp}). Methods in the first group constrain the geometry deformation only to the surface (e.g.,~\cite{lapedit,variational,arap}), while methods in the second group deform the domain that contains the surface (e.g.,~\cite{primo, Sumner2007, biharmonic}). More recently, neural mesh deformation approaches started gaining popularity (e.g.,~\cite{neural_cages}). A~meshless method for modeling deformations and creating animations was addressed by Adams et al. \cite{Adams2008}. They proposed a~method that finds a~smooth deformation field based on prescribed deformations in a~set of discrete points (handles), matches the prescribed deformations, and is additionally as rigid and volume-reserving as possible. This method allows not only the deformation from the source shape to the target shape but also the animation of this deformation. Furthermore, the deformation is topology-aware and can, therefore, distinguish geodesically distant parts of the shape and allow the insertion of obstacles into the deformation field. However, the method does not work with original point trajectories since the input shape is static. Static editing methods not only support the creative process in design and animation but also serve as a~basis for other methods in various different scenarios, e.g., correspondence estimation~\cite{divfree}, tracking~\cite{bojsen2012tracking}, RGB-D reconstruction~\cite{killingfusion}, or compression~\cite{doumanoglou}. For a~more elaborate discussion of the state of the art in this field, we refer the reader to the survey of Yuan et al.~\cite{edit_survey}. In order to edit a~mesh sequence representing a~moving object, another challenge must be addressed: even small discrepancies in the edited geometry between consecutive frames can be easily spotted by an observer~\cite{STED}. Thus, the goal is to make the editing process temporally coherent, which cannot be automatically achieved without considering inter-frame correspondences.

Some progress has already been made in the case of editing dynamic (animated) meshes, a~special type of mesh sequences, in which all the frames share a~common connectivity. This property explicitly implies global bijective inter-frame correspondence of vertices. Dynamic meshes are almost exclusively a~result of the continuous deformation of a~static mesh template. If the deformation parameters are known, some methods, especially those deforming the domain around the surface, permit slight edits to the initial template. In the case of skinned meshes, a~lot of work has been done on applying changes to evolving skeleton poses (e.g.,~\cite{gleicher97, kim2007motion, wu2008space, skeleton_laplacian, animesh, optimo, magic_wand}).

In the case of an unknown deformation model, it might be possible to approximate the sequence by a~similar one represented by skinning~\cite{dmskinning,aguiarskinning,rigging} or cage-based deformation~\cite{cager}. This, however, cannot be performed on non-articulated motions (e.g., cloth simulations)~\cite{kircher_editing}. To this end, several methods were proposed as extensions of the static approaches with positional constraints or vertex displacements smoothly blended between frames~\cite{kircher_editing, xu_gradient, kircher_free_form, tejera1, aguiar, tejera2, zhao, abrevaya}. Cashman and Hormann~\cite{cashman} represent the sequence as a~curve in a~learned pose space. Not only does this approach allow creating loops, temporal resampling, and motion transfer, but it is also able to realize slight edits by modifying the curve. Additionally, some methods interpret the original sequence as rest poses of a~deformable object and perform the edits via an elastic model simulation with damping forces~\cite{barbic,hildebrandt,lisimulation1,lisimulation2,liysimulation}.

Using the aforementioned approaches for editing general mesh sequences with varying connectivity is not feasible since, in general, for these sequences, there is no known deformation model, such as a~skeleton, that is also time-consistent for the entire sequence. Although in some specific cases, such as human animations, the skeleton of the sequence can be estimated, for general data, these approaches are not applicable since the actual topology of the edited surface is often unknown, and the observed topology may vary over the sequence due to surface self-contact.

To the best of our knowledge, there is only a~single editing framework suitable for temporally coherent mesh sequences with varying connectivity~\cite{tvmeditingyang}. The temporal coherence of edits is achieved by smoothing between adjacent frames utilizing estimated frame-to-frame correspondences. Unfortunately, the method allows only the enhancement of details.

Another problem related to sequence editing is the construction of motion graphs. The objective in this scenario is to identify visually similar frames at different time points in the sequence, between which intermediate frames are inserted in order to create a~smooth transition. One can then utilize these transitions to create a~new sequence with a~different order of motions. Motion graphs have been applied to motion capture data~\cite{heck,kovar}, dynamic meshes~\cite{casas1,casas2,casas3,huang_graphs}, and even general mesh sequences~\cite{Xu2008,prada}, however, surface correspondences must be established between pairs of similar frames.

\begin{figure}[ht]
  \centering
  \includegraphics[scale=0.35]{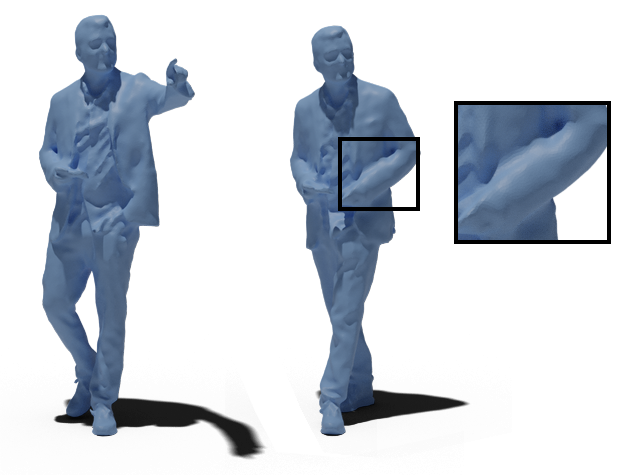}
  \caption{Example of the disappearance of part of the surface due to self-contact. Left: The entire surface of the left hand is visible. Right: Part of the surface of the left hand is missing.}
  \label{fig:hidden_surface}
\end{figure}

In order to achieve more global edits in both space and time, it is necessary either to convert the general mesh sequence to a~dynamic mesh~\cite{tevs}, which is often hard or impossible due to a~lack of correspondence bijectivity caused by self-contact, or to construct a~robust temporal model, which captures the global correspondences (see Fig.~\ref{fig:hidden_surface}). Bojsen-Hansen et al.~\cite{bojsen2012tracking} proposed a~tracking method based on embedded deformations, which was able to adapt to changing topology. As a~result, they are able to coherently map various functions (e.g., colors or displacements) to the evolving surface. However, they assume bijectivity of the correspondences, with the only exception occurring when the topology of the surfaces changes. A~volumetric approach is more suitable since the correspondences of the volume enclosed by the surface are more likely to be bijective. Huang et al.~\cite{huang_tracking} used non-rigid registration of centroidal Voronoi tessellations. Dvořák et al.~\cite{Dvorak2021} additionally employed optimization, which enforced a~more fluent motion of the tracked volume elements. Both these approaches are, however, prone to assignment errors, which, if incorporated into a~deformation model, would result in high-frequency artifacts. In their subsequent work, Dvořák et al.~\cite{Dvorak2022, Dvorak2023} were able to significantly reduce the tracking errors by means of additional improvements in the optimization.

\section{Editing}
The input of the method is a set of $F$ frames, each consisting of a set of vertices $\mathbf{v}_{i,f}, i=1..V_f, f=1..F$ (triplets of coordinates) and triangles $\mathbf{t}_{i,f}, i=1..T_f, f=1..F$ (triplets of vertex indices).  To obtain a~temporally consistent deformation of a~TVM, our method exploits the volume tracking method proposed by Dvořák et al.~\cite{Dvorak2022}. Its result is a~set of volume elements represented by their centroids (denoted \emph{centers}) $\mathcal{C} = \left \{ \mathbf{c}_1, \mathbf{c}_2, \ldots, \mathbf{c}_n \right \}$, where $\mathbf{c}_i = \left [ \mathbf{x}_{i, 1}, \mathbf{x}_{i, 2}, \ldots, \mathbf{x}_{i, F} \right ] \in \mathbb{R}^{3F}$ is a~trajectory of the $i$-th center consisting of its positions $\mathbf{x}_{i, j}$ in each frame $j$. 

The ARAP volume tracking method used to estimate temporal correspondences is based on estimating the motion of volume elements (centers) that uniformly cover the tracked volume. The initialization of the method is done by uniformly distributing the centers within the volume of the first frame of the sequence. Next, the set of centers is propagated from each frame to the next, while in each frame, the positions of the centers are locally updated by optimizing an energy function that penalizes uneven distribution of centers in the covered volume and deviation from the rigidity of motion for centers that are considered connected (i.e., lie nearby and move similarly in the past frames). The correspondence between the tracked centers is then already given implicitly, and there is no need to match them with each other. Note that the proposed method is not dependent on the implementation of this particular tracking method, and any other method for estimating discrete volume correspondences can be used instead. Alternatively, dense volume correspondences can be used as well, which can be transformed into discrete ones by uniform sampling. In the rest of the paper, we use a bar to denote an edited version of a~center or a~vertex, i.e., $\overline{\mathbf{x}}_{i, j}$ denotes the edited version of a center position $\mathbf{x}_{i, j}$.

Our TVM editing method consists of three steps: editing of tracked centers in one frame, the subsequent propagation of this deformation to centers of other TVM frames, and the final deformation of the surface in all frames. This procedure provides sufficient freedom to change the surface shape in many practical scenarios. 

The deformation model can be decomposed into three functions: $\mathcal{D}_{\text{intra}}$, $\mathcal{D}_{\text{inter}}$, and $\mathcal{D}_{\text{surface}}$. First, the user chooses a frame $f$, in which they perform the edit by specifying transformations $\left\{\mathbf{T}_{j,f}\right\}_{j\in{E}}$, for a chosen set $E$ of centers. Next, the function $\mathcal{D}_{\text{intra}}$ assigns distributed transformations $\mathbf{T}_{i, f}, i\notin E$ to the remaining centers in frame $f$, based on the user input and their original positions $\mathbf{x}_{i,f}$:
\begin{equation}
\label{eq:eq02a}
\mathbf{T}_{i,f}=\mathcal{D}_{\text{intra}}\left(i,f,\mathcal{C},\{\mathbf{T}_{j,f}\}_{j\in E},E\right).
\end{equation}

Note that the function receives the complete trajectories of all the centers $\mathcal{C}$, and through this data, it can potentially exploit information from the whole sequence even when distributing the transformations within a single frame. The transformations could be described by translation vectors between the original center position $\mathbf{x}_{j, f}$ and the imposed center position $\overline{\mathbf{x}}_{j,f}$ as $\mathbf{T}_j=\overline{\mathbf{x}}_{j,f}-\mathbf{x}_{j,f}$. However, in order to build a~more natural deformation model, it is beneficial (as demonstrated later in Section \ref{sec:experiments}) to describe the transformation of the centers not as a~translation, but as a~rigid transformation that also contains a~rotation component. 

A suitable representation of rigid transformations, which is known to provide the possibility of natural-looking blending~\cite{Kavan2008}, is the representation by unit dual quaternions. Using a~dual quaternion $\mathbf{A}=A+\epsilon B$, where $\epsilon$ is a~dual unit ($\epsilon^2=0$), it is possible to express a~rotation given by a~quaternion $R$ using a~dual quaternion $\mathbf{A}_r$ as $\mathbf{A}_r=R+\mathbf{0}\epsilon$. Unlike quaternions, dual quaternions can also describe a~translation by a~vector $\mathbf{t}=(t_x,t_y,t_z)$ as $\mathbf{A}_t=\mathbf{0}+\left(\frac{t_x}{2},\frac{t_y}{2},\frac{t_z}{2},0\right)\epsilon$. However, not all dual quaternions represent a~\emph{rigid} transformation. Only unit dual quaternions with $\|\mathbf{A}\|=1$ satisfy this property. Thus, the blending of dual quaternions uses normalization of the resulting dual quaternion to project back into the space of rigid transformations. In the rest of the paper, we use capital bold letter $\mathbf{A}$ to denote general dual quaterion of non-unit length, the hat symbol $\,\hat{}\,$ to denote a unit norm dual quaternion $\hat{\mathbf{A}}$ and the star symbol~$^*$ to denote a~conjugate dual quaternion $\mathbf{A}^*$. With an appropriate representation of a transformation associated with each center, the new positions of centers $\overline{\mathbf{x}}_{i,f}$ can be expressed as $\overline{\mathbf{x}}_{i,f}=\mathbf{T}_{i,f}\left(\mathbf{x}_{i,f}\right)$.


The advantage of unit dual quaternions as a representation of rigid transformations is their inherent ability to compose transformations by multiplying them (similar to transformation matrices). However, unlike transformation matrices, it is not necessary to search for the inverse of a matrix to find the inverse of a transformation. Moreover, when blending unit dual quaternions, which results in a quaternion that no longer satisfies the unit norm, it is simply possible to project this quaternion back into the space of rigid transformations by normalizing it. The same operation in the case of transformation matrices requires finding an orthogonal matrix that is somehow close to the bled transformation.

A function $\mathcal{D}_{\text{inter}}$ assigns the propagated transformation $\mathbf{T}_{i, g}$ to the tracked centers in each frame $g \neq f$, based on the original center positions $\{\mathbf{x}_{j, g}\}_{j=1}^n$, their corresponding positions in the edited frame $\{\mathbf{x}_{j, f}\}_{j=1}^n$, and a transformation in the edited frame $\mathbf{T}_{i, f}$: 
\begin{equation*}
\label{eq:eq02b}
\mathbf{T}_{i, g}=\mathcal{D}_{\text{inter}}(i, g, \{\mathbf{x}_{j, g}\}_{j=1}^n,\{\mathbf{x}_{j, f}\}_{j=1}^n,\mathbf{T}_{i, f}).
\end{equation*}

The output of this function consists of transformations that, when applied to the original positions of the centers, give the positions of the centers after deformation, preserving the original trajectories as much as possible and, at the same time, reflecting the change in shape with respect to the edited frame.

Finally, a function $\mathcal{D}_{\text{surface}}$ assigns new mesh vertex positions $\overline{\mathbf{v}}_{i,f}$ based on the original positions $\mathbf{v}_{i,f}$, the original positions of centers $\{\mathbf{x}_{j,f}\}_{i=1}^n$, and the center transformations $\{\mathbf{T}_{j,f}\}_{i=1}^n$, as follows: 
\begin{equation*}
\label{eq:eq02c}
\overline{\mathbf{v}}_{i,f}=\mathcal{D}_{\text{surface}}(i, f, \mathbf{v}_{i,f},\{\mathbf{x}_{j,f}\}_{i=1}^n,\{\mathbf{T}_{j,f}\}_{i=1}^n).
\end{equation*}
\section{Deformation model}\label{sec:deformation-model}
Our framework allows editing a~TVM's geometry using three operations available to the user. 
The supported operations are as follows:
\begin{itemize}
    \item the rigid transformation of a~set of centers,
    \item center-based inflation and deflation, and
    \item closing the sequence into a~loop.
\end{itemize}

\subsection{Intra-frame deformation of centers}
To propagate the deformation within the frame while respecting the true topology of the data, the distance between the centers should correspond to the distance in the occupied volume, rather than the distance in space, so that centers that are located close to each other in space but each belongs to a~different component or are in a~geodesically distant part of the same component do not affect each other.

In order to propagate the deformation of the selected center to the other centers within a~single frame, taking into account the true global topology of the surface, we calculate the \emph{affinity} between individual centers $a(i,j)\in[0,1]$, which measures how tightly the two centers are bound together. Affinity calculation depends on the maximum distance between the centers across the entire sequence. The farther apart the centers get during the sequence, the lower the affinity between them will be. The resulting affinity is then given by a~Gaussian radial basis function of the maximum distance, with a~shape parameter $\alpha$:
\begin{equation*}
\label{eq:eq03a}
a(i,j)=exp\left(-(\alpha \max_{f}(\|\mathbf{x}_{i,f}-\mathbf{x}_{j,f}\|))^2\right).
\end{equation*}

Radial basis functions, such as the Gaussian function, are naturally a suitable apparatus for modeling a function in the neighborhood of a given center whose value increases with distance from the center. They are, therefore, a suitable tool for representing, for example, the affinity between centers.

The suitable value of the $\alpha$ parameter depends on the global scale of the data. The $\alpha$ parameter affects the rate at which the affinity of the centers decreases as their maximum distance increases (double $\alpha$ means that the affinity reaches the same value at half the squared distance). In our experiments, we chose parameter values in the range of $10^-2$ to $10^1$.

\begin{figure}[ht]
  \centering
  \includegraphics[scale=0.1]{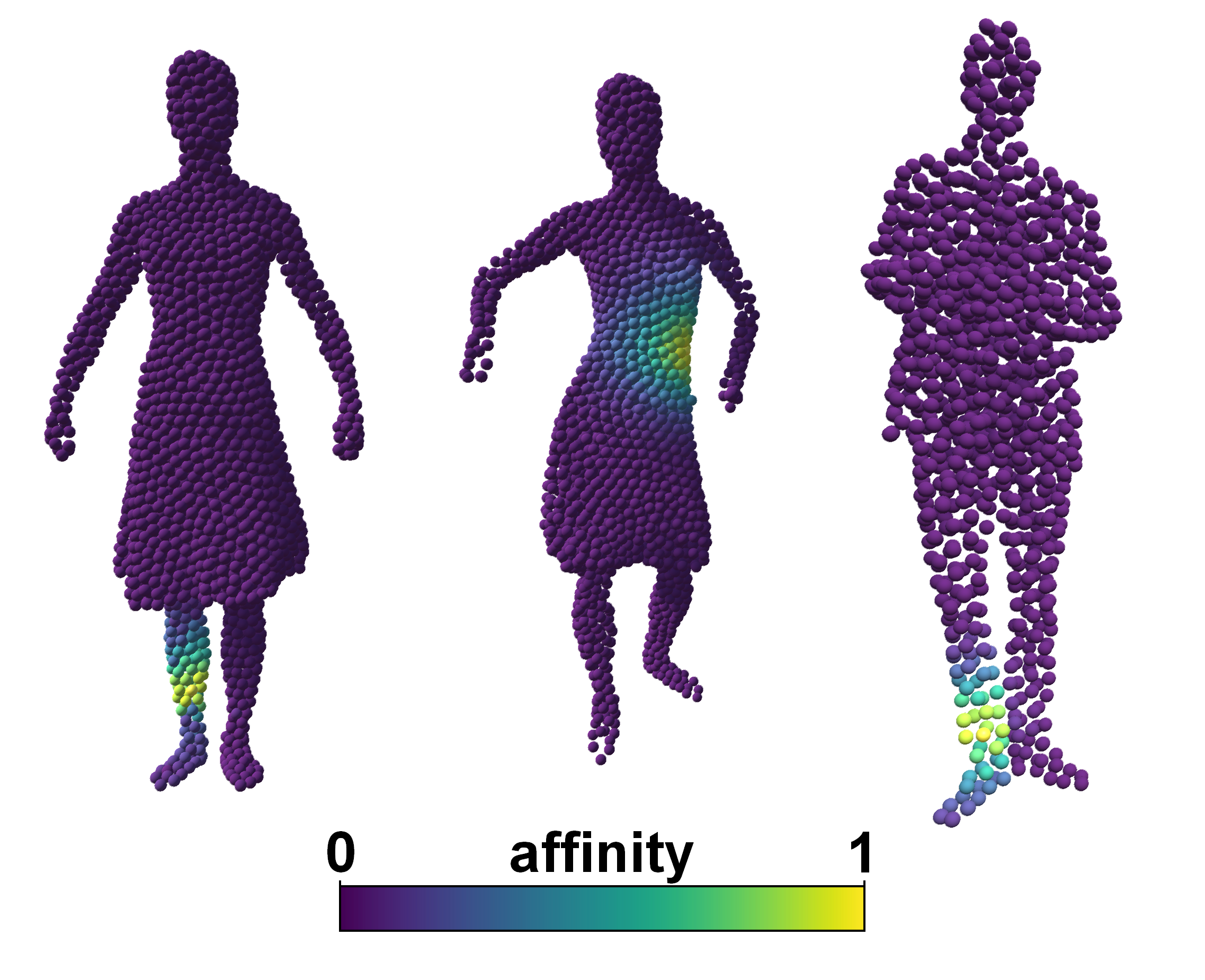}
  \caption{Visualization of the affinity of centers. Spatially close centers that are geodesically distant have a~low affinity.}
  \label{fig:affinity}
\end{figure}

These affinities prevent parts of the surface that are close in the sense of the Euclidean distance, but that are geodesically distant, from affecting each other (see Fig. \ref{fig:affinity}), as long as they get far enough apart at some point during the sequence. Therefore, the affinity expressed in this way naturally captures information from all the frames of the sequence. 

With the calculated affinity and the set $E$ of edited centers, where the center $\mathbf{x}_{i,f}$ is transformed by a~user-defined rigid transformation described by a~unit dual quaternion $\hat{\mathbf{T}}_i$, the blended transformations of the other centers can be expressed as:
\begin{equation}
\label{eq:eq03b}
\mathbf{T}=\frac{1}{\sum_{e\in E} a(i,e)}\sum_{e\in E} a(i,e) \hat{\mathbf{T}}_e.
\end{equation}

To project the blended dual quaternion back to the space of rigid transformations, it is necessary to normalize it. The final distributed transformations $\hat{\mathbf{T}}_i$ are therefore given by
\begin{equation*}
\label{eq:eq03bb}
\hat{\mathit{T}}_i=\frac{\mathbf{T}_i}{\|\mathbf{T}_i\|}.
\end{equation*}

To demonstrate the advantages of using an affinity-dependent maximum center distance, we compare this method with a~naïve method, in which the argument of the radial basis function is the Euclidean distance between centers in the current frame. A~comparison of these methods is provided in Section \ref{sec:experiments} (see Fig. \ref{fig:samba_affinities}).

\subsection{Inter-frame deformation of centers}\label{sec:inter}
After distributing the deformation to all centers of a~single frame, it is necessary to propagate this deformation to the centers of the remaining frames of the sequence. Even though the temporal correspondence of centers is known, the deformations of individual centers must be adjusted in order to take into account the motion that occurs between the source and the target frame. The edits are typically local, and it is therefore necessary to find a~local coordinate frame in which the transformations should be applied. To this end, it is possible to use the Kabsch algorithm \cite{Kabsch1976} to find the optimal transformations of centers in the remaining frames. Specifically, for each center, we calculate the optimal rigid transformation using the $K$ centers with the greatest affinity to it. We use the weighted Kabsch algorithm\cite{SorkineRabinovich:SVD-rotations:2016}, in which the weights of the individual centers are equal to the affinity between them.

For each center $\mathbf{x}_{i,f}$, the algorithm calculates the origin of the center's neighborhood $\mathbf{o}_i$ in the source frame $f$, in which the deformation was performed, and in the target frame $g$, to which the deformation is propagated to, using:
\begin{equation*}
\label{eq:eq03c}
\mathbf{o}_i=\frac{\sum_{j=1}^K a(i,j) \mathbf{x}_{i,f}}{\sum_{j=1}^K a(i,j)}.
\end{equation*}

Next, we use the Kabsch algorithm with a~matrix of weights $\mathbb{W}_i=diag(\left[a(i,j)\right]_{j=1}^K)$ to compute the optimal rotation between centers of the source and the target frame. The optimal rotation can be expressed by a~dual quaternion $\hat{R}_i$, and the translation to the origin of the neighborhood of the centers $\mathbf{o}_i$ can be expressed by a~dual quaternion $\hat{\mathbf{O}}_i$.

It is now possible to propagate the deformation $\hat{\mathbf{T}}_{i,f}$ from the source frame $f$ to a~target frame $g$. A center $\mathbf{x}_{i,g}$ in a~target frame is first translated to the origin using $\hat{\mathbf{O}}_{i,f}^*$. Next, the inverse rotation $\hat{\mathbf{R}^*}_i$ is applied. After that, the translation to the centroid of the center's neighborhood in a~source frame is performed $\hat{\mathbf{O}}_{i,f}$, followed by the deformation $\hat{\mathbf{T}}_{i,f}$ itself. The center is then translated back to the origin using $\hat{\mathbf{O}
}_{i,f}^*$, rotated by $\hat{\mathbf{R}}_i$, and translated back to the centroid of the center's neighborhood in the target frame using $\hat{\mathbf{O}}_{i,g}$. The composition of the above dual quaternions transforms the center from its position and orientation in the target frame $g$ to its position and orientation in frame $f$. Here, the deformation is performed, and then the center is transformed back. This entire series of transformations can be expressed as the product of the individual dual quaternions, by which the original center $\mathbf{x}_{i,g}$ is subsequently transformed:
\begin{equation*}
\label{eq:eq03cc}
\overline{\mathbf{x}}_{i,g}= \hat{\mathbf{O}}_{i,g} \hat{\mathbf{R}}_i \hat{\mathbf{O}}_{i,f}^* \hat{\mathbf{T}}_i \hat{\mathbf{O}}_{i,f} \hat{\mathbf{R}}_i^* \hat{\mathbf{O}}_{i,g}^* (\mathbf{x}_{i,g}).
\end{equation*}

To allow for editing that is local in terms of the time axis with sufficient smoothness without causing temporal artifacts, the framework allows blending the propagated transformation of a~center with the identity transformation $\hat{e}$ with a~time-dependent parameter $\varphi(t)\in[0,1]$ (\emph{time decay}). The interpolation result is then again normalized to restore the unit size of a dual quaternion, as follows:
\begin{equation*}
\label{eq:eq03decay}
\mathbf{T}_i=\frac{(1-\varphi)\hat{\mathbf{T}}_i + \varphi \hat{e}}{\|(1-\varphi)\hat{\mathbf{T}}_i + \varphi \hat{e}\|}.
\end{equation*}
For example, for linear time decay $\varphi(t)=max(0, r-|t-t_0|)$, at times $t=t_0-r$ and $t=t_0+r$, the propagated motion is propagated with weight $0$, i.e., the trajectory of the sequence does not change. Conversely, for $t=t_0$, the motion is propagated with weight $1$, and the resulting trajectory is determined by the composition of the original center transformation and the applied deformation. More complex functions $\varphi(t)$ with larger continuity order can be used.

\subsection{Surface deformation}
Once we have the original positions of the centers $\{\mathbf{x}_{i,f}\}_{i=1}^n$ and the distributed transformations of the centers after deformation $\hat{T}_{i,f}$, we deform the surface by calculating the new positions of the vertices of the triangle meshes. Since it is desirable that the deformation model sufficiently reflects the natural properties of elastic bodies in the real world, it aims to satisfy the following properties:


\begin{itemize}
    \item Deforming a~vertex $\mathbf{v}$ with a~neighborhood of centers $\{\mathbf{x}_{i,f}\}_{i=1}^K$, the resulting deformation is given by the interpolation of the deformations in the centers $\mathbf{x}_{i,f}$ with weights $w_i\in [0,1]$, the sum of these weights being equal to one. In other words, the deformation at point $\mathbf{x}$ is given by a~convex combination of the deformations of the surrounding centers.
    \item The function $w_i(\mathbf{v}): \mathbb{R}^3\rightarrow\mathbb{R}$, assigning deformation weights to the volume centers $\mathbf{x}_{i,f}$, is continuous on the surface.
    \item The points $\mathbf{v}$ of the surface where the function $w_i(\mathbf{v}) > 0$ form a~compact space. Thus, the effect of the center with respect to the surface deformation is local.
    \item For points whose position is the same as the position of the $j$-th center, the weight $w_j$ is equal to one, and the other weights are zero: 
    \begin{equation*}
    \label{eq:eq03q}
    w_j(\mathbf{x}_{i,f})=\left\{ \begin{matrix}
        1\text{ for } j=i \\
        0\text{ for } j\neq i
    \end{matrix} \right..
    \end{equation*}
    
\end{itemize}

One possible way to deform the mesh geometry based on the difference in center positions is to use the Embedded Deformations method~\cite{Sumner2007}, which consists of finding the $k$-nearest centers, calculating their weights based on their distance from the deformed vertex, and weighting the sum of the translation vectors at the nearest centers. Having $k$-nearest centers $\{\mathbf{x}_{i,f}\}_{i=1}^k$ ordered by their distance from the vertex $\mathbf{v}$ ($\mathbf{x}_{k,f}$ is the farthest center of the $k$-nearest centers), the weights $w_i$ are given as
\begin{equation*}
w_i = \left(1-\frac{\|\mathbf{v}-\mathbf{x}_{i,f}\|}{\|\mathbf{v}-\mathbf{x}_{k,f}\|}\right)^2.
\end{equation*}

Weights $w_i$ can be used to compute the interpolated transformation of the vertex. The interpolated transformation is given by
\begin{equation*}
\label{eq:eq03i}
\mathbf{T}_v=\frac{\sum_{i=1}^k w_i \hat{\mathbf{T}}_i}{\sum_{i=1}^k w_i}.
\end{equation*}

The vertex position $\mathbf{v}$ is then transformed by the normalized dual quaternion $\mathbf{T}_v$:
\begin{equation*}
\label{eq:eq03ii}
\overline{\mathbf{v}}=\frac{\mathbf{T}_v}{\|\mathbf{T}_v\|}\left(\mathbf{v}\right).
\end{equation*}

The disadvantage of this approach is that the area of influence of a center is fully determined by the spatial distance and does not respect the topology of the edited surface. When deforming the~surface, the set of k-nearest centers may contain centers that are close in terms of spatial distance, but their distance on the surface or distance in the volume can be much larger or even infinite (in case they belong to disconnected components).

Furthermore, the method of Embedded Deformations does not perform interpolation, i.e. at the position of a~center $\mathbf{x}_{i,f}$, its weight $w_i$ is not equal to one, and the other weights are not zero, which in combination with continuity of deformation is a~crucial property for keeping the centers inside the volume during deformation.

These properties can also be achieved by using Bounded Biharmonic Weights (\cite{biharmonic}), which have better properties in terms of locality and sparsity. However, this advantage becomes less pronounced when a~sufficient number of volume centers are used. The disadvantage of Bounded Biharmonic Weights is that they require an optimization problem to be solved, while for practical cases with thousands of centers and hundreds of thousands of vertices in several hundreds of frames, a~closed form solution is preferable.

In order to address the problems of the above-sketched approaches, we propose an alternative way to compute the blending weights. The proposed weights achieve locality by normalizing the weight of the most distant center to zero, similarly to weights of Embedded Deformations. However, our weights also \emph{incorporate the affinity of the centers} and satisfy the interpolation condition. When editing a~vertex $\mathbf{v}$, we first find the nearest center $\mathbf{x}_{1,f}$ in frame $f$ as \begin{equation*}
\label{eq:eq03l}
\mathbf{x}_{1,f}=\arg \min_{\mathbf{x}_{i,f}} \|\mathbf{x}_{i,f}-\mathbf{v}\|.
\end{equation*}
This step assumes that the density of centers in the volume and the accuracy of the volume-tracking is sufficient for the closest center to best approximate the trajectory of the point on the surface given by the vertex.

The algorithm then proceeds with finding the next $k-1$ centers, but no longer based on the Euclidean distance to the vertex, but as the $k-1$ \emph{most affine} centers to center $\mathbf{x}_{1,f}$. This ensures that centers from topologically distant parts of the volume do not appear in the set. The center $\mathbf{x}_{1,f}$ and the $k-1$ centers of its most affine neighbors form the set $\mathcal{C}_k$. Next, we calculate the distances $d_i$ between the centers of the set $\mathcal{C}_k$ and the vertex $\mathbf{v}$ as
\begin{equation*}
\label{eq:eq03m}
d_i=\|\mathbf{x}_{i,f}-\mathbf{v}\|, i\in\{1, 2, \cdots, k\}.
\end{equation*}

Based on the distances $d_i$, it is possible to calculate a~set of pseudo-weights $w'$ as 
\begin{equation*}
\label{eq:eq03n}
w'_i=\lim_{\delta\rightarrow d_1+} \frac{\exp(\frac{-d_i}{\sigma \delta})}{\sum_{j=1}^k\exp(\frac{-d_j}{\sigma \delta})}.
\end{equation*}
These pseudo-weights satisfy the interpolation condition, i.e., if a~vertex $\mathbf{v}$ spatially coincides with its nearest center $\mathbf{x}_{1,f}$, then its pseudo-weight will be equal to one, and the weights of the other centers will be zero. This function is based on a \textit{smooth minimum} function (Boltzmann operator) \cite{Boltzmann1868}, which continuously divides the value of one between the centers so that the nearby centers get more weight than the more distant ones. To make the expression easier to evaluate, we replace the limit by adding a~sufficiently small $\epsilon$ in the denominator of the argument:
\begin{equation*}
\label{eq:eq03o}
w'_i\approx\frac{\exp(\frac{-d_i}{\sigma d_1+\epsilon})}{\sum_{j=1}^k\exp(\frac{-d_j}{\sigma d_1+\epsilon})}.
\end{equation*}
The $\sigma$ parameter controls the smoothness of the weights. The value of the $\sigma$ parameter determines the proportion with which closer centers affect resulting deformation compared to more distant ones when blending deformation centers. Lower values of the $\sigma$ parameter indicate a more similar influence of closer and more distant centers. The value of the $\sigma$ parameter is a scale-invariant. In our experiments, we used $\sigma$=2. Finally, the pseudo-weights $w'_i$ are normalized so that the weight of the most distant center $c_k$ is zero using the following formula and then normalized so that their sum is equal to one:
\begin{equation*}
w_i=\frac{w'_i-\min_{j}(w'_j)}{\sum_k \left(w'_k-\min_{j}(w'_j)\right)}.
\end{equation*} 

\subsection{Surface subdivision}
Since the final deformation of the surface represented by the triangle mesh is not rigid, stretching or shrinking of the surface occurs when this deformation is applied. In the case of surface stretching, by which the area of the deformed triangles increases, the deformed mesh may suffer from artifacts in the form of plateaus caused by triangles that are too large to accurately approximate the smooth, deformed surface. 

To address this problem, it is possible to extend the surface deformation model to detect those cases in which the tessellation of the triangle mesh is too coarse with respect to the given deformation and to adaptively subdivide it in such cases.

For the given original vertex positions $\mathbf{v}_i$ and the deformed vertex positions $\overline{\mathbf{v}}_i$, the algorithm checks each edge $(\mathbf{v}_i,\mathbf{v}_j)$. If, for a~given edge, the rotation component of the difference in the transformation of its vertices has a~large angle, then the algorithm splits the edge in order to prevent the artifacts caused by insufficient resolution of the tessellation of the surface. For transformations $\hat{\mathbf{T}}_i$ and $\hat{\mathbf{T}}_j$, the difference in transformation is expressed as a~product $\hat{\mathbf{T}}_{\Delta}=\hat{\mathbf{T}}_i\hat{\mathbf{T}}_j^*$. The rotation component is represented by the real component $A$ of the dual quaternion $\hat{\mathbf{T}}_{\Delta}=A+B\epsilon$. If the real component $w_a$ of this quaternion $A=\left(x,y,z,w\right)$, which describes the cosine of the angle of the given rotation, is lower than a~given threshold $\zeta_{min}$ (in our case, $\zeta_{min}=0.1\pi$), then the deformation described by the transformations $\hat{\mathbf{T}}_i$ and $\hat{\mathbf{T}}_j$ causes too much bending of the edge, and thus it should be split. Note that the splitting could alternatively be derived from the stretching of the edge length caused by the editing. However, stretching alone is not a~good indicator since edges, although stretched, often remain straight, and in such cases, inserting a~new vertex is superfluous.

\begin{figure}[ht]
  \centering
  \includegraphics[scale=0.125]{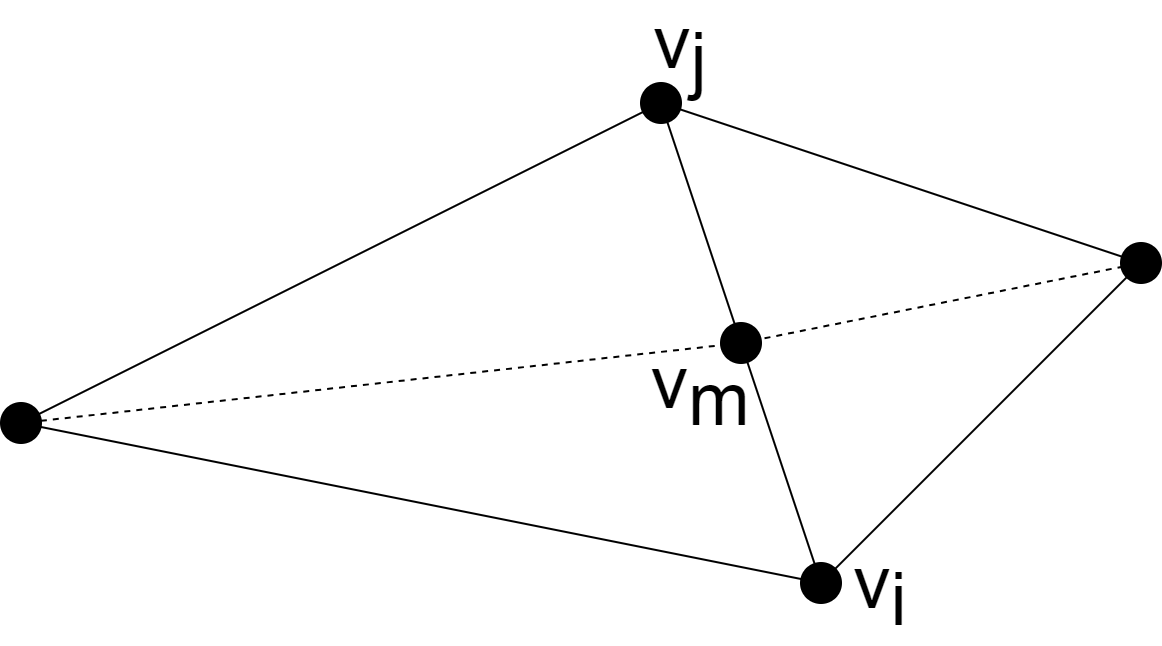}
  \caption{Diagram illustrating the subdivision of an edge and its incident triangle faces.}
  \label{fig:edge_sub}
\end{figure}

A new vertex $\mathbf{v}_m$ is inserted in the middle of the original edge, and the incident triangle faces are subdivided (see Fig.~\ref{fig:edge_sub}). Weights are then calculated for the new vertex $\mathbf{v}_m$ to compute the deformation, and the vertex position is deformed in the same way as the original mesh vertices. The subdivision process is stopped when there is no edge to split or when the maximum number of iterations is reached.

\subsection{Rigid motions}
One of the properties that the surface deformation model should exhibit is that if a~global rigid motion is applied to the centers, the deformation model should apply the same rigid motion to the vertices of the deformed mesh. This property is known as rigid motion invariance.

In the case of the deformation of the centers of one frame described by Eq.~\ref{eq:eq03b}, it can easily be shown that when the same $\hat{\mathbf{T}}_{rigid}$ transformation is applied to all effectors, this exact transformation is propagated to the other centers:
\begin{equation*}
\label{eq:eq03ra}
\mathbf{T}_i=\frac{1}{\sum_e a(i,e)}\sum_e a(i,e) \hat{\mathbf{T}}_{rigid}=\frac{\sum_e a(i,e)}{\sum_e a(i,e)}\hat{\mathbf{T}}_{rigid}=\hat{\mathbf{T}}_{rigid}.
\end{equation*}

Similarly, rigid motion invariance can be shown for the deformation of the mesh vertices. Since each vertex is transformed by a~convex combination of rigid transformations, which are all identical in this case, the resulting vertex transformation is also the same. Finally, it can also be shown that if frames differ only by a~rigid transformation, then all the local coordinate systems are estimated equal by the Kabsch algorithm, and the resulting effect is a~composition of the imposed rigid transformation and the rigid movement that occurs in each frame.

\section{Other applications}

\subsection{Inflation/deflation}
In addition to the editing done manually by changing the positions of a~selected set of centers, our framework also allows the user to inflate and deflate the volume using the \emph{center attraction} tool. The procedure is that the user selects a~center $\mathbf{x}_{i,f}$, around which the edit is done, and a~parameter $\beta$, which determines how much the volume should be locally enlarged or reduced. 

Subsequently, the positions of the surrounding centers $\mathbf{x}_{j,f}$ are changed by translating them in the direction given by the difference between the position of center $\mathbf{x}_{j,f}$ and the position of the selected center $\mathbf{x}_{i,f}$. The length of the translation vector is directly proportional to the affinity of the centers $a(i, j)$ and the beta parameter:
\begin{equation*}
\label{eq:eq03k}
\hat{\mathbf{T}}_j=\beta \cdot a(i,j) (\mathbf{x}_{j,f} - \mathbf{x}_{i,f})\epsilon.
\end{equation*}

\begin{figure}[ht]
  \centering
  \includegraphics[scale=0.25]{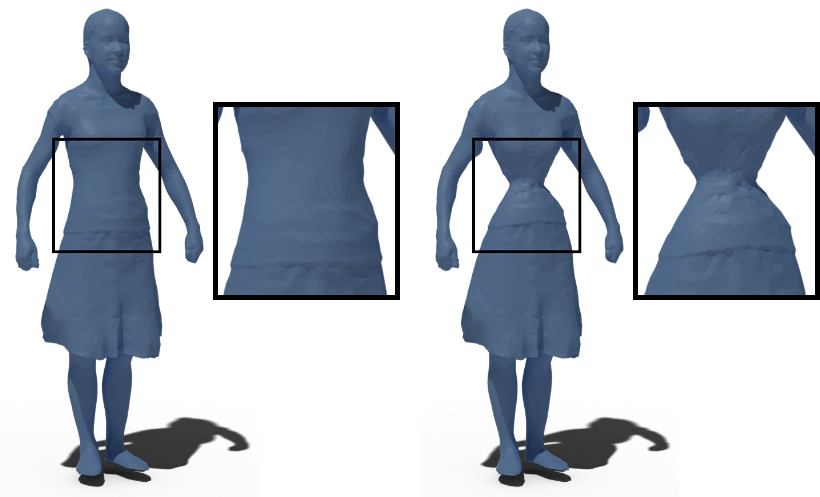}
  \caption{Example of increasing center attraction to slim the figure at the waist. Left: Original. Right: Edited. Note that because center affinity has been used for distributing the edit, the hands of the model remain unaffected.}
  \label{fig:att}
\end{figure}

\begin{figure}[ht]
  \centering
  \includegraphics[scale=0.12]{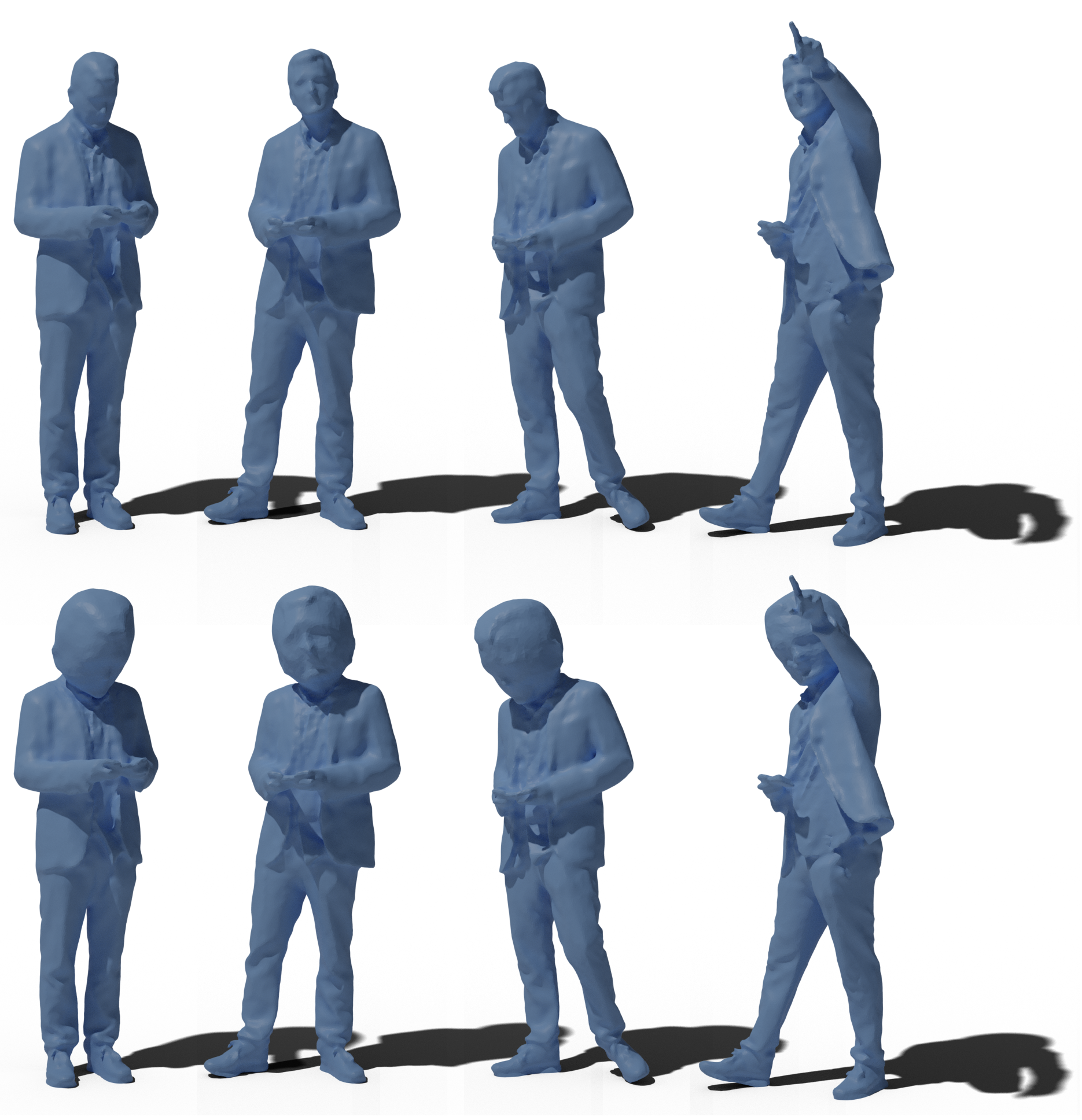}
  \caption{Example of inflation of centers used to increase head size. Top: Original sequence. Bottom: Edited sequence.}
  \label{fig:dis}
\end{figure}

The results of using this editing tool are shown in Figs. \ref{fig:att} and~\ref{fig:dis}.

\subsection{Closing loops}
In some applications, it is required that a~time-varying mesh can be continuously looped without visible temporal artifacts between the last and first frames. However, acquiring such data by methods based on the 3D reconstruction of an actual scene is challenging. In the case of human motion tracking, to achieve a~smooth transition between the last and the first frame, the tracked individuals must be in exactly the same pose at the end of the sequence as they were at the beginning of the sequence. For this reason, post-processing of the captured sequence is often required to suppress the temporal artifact. 

Our framework offers two options for looping a~sequence. The first option is a~fully automatic approach based on computing the deformation between the first and the last frames of the sequence and then propagating the deformation across the sequence. However, this approach requires high accuracy and a~high degree of consistency of the tracked centers. In the case in which these conditions are not met, a~semi-automatic method can be used, in which the first frame of the sequence is manually edited by the user using the available operations to match the last frame of the sequence as closely as possible (or vice versa), and this deformation is then automatically propagated with time decay. 

As with the propagation of deformations within a~sequence, to find the transformations between the last and the first frame, the Kabsch algorithm is used. For each center $\mathbf{x}_{i,f}$, a~translation vector between the origin and the weighted centroid of the set of its most affine centers is found ($\hat{\mathbf{O}}_{i,1}$ in the first frame and $\hat{\mathbf{O}}_{i,n}$ in the last frame), together with the optimal rotation $\hat{\mathbf{R}}_i$. The resulting transformation from the last frame of the sequence to the first frame is then given by the composition of these transformations:
\begin{equation*}
\label{eq:eq03kk}
\hat{\mathbf{T}}_i=\hat{\mathbf{O}}_{i,n}^*\hat{\mathbf{R}}_i\hat{\mathbf{O}}_{i,1}.
\end{equation*}

The last frame of the sequence is deformed by applying these transformations, and this deformation is propagated with time decay to the previous frames as described in Section \ref{sec:inter}.



\subsection{Vertex editing}
\begin{figure}[ht]
  \centering
  \includegraphics[width=0.8\linewidth]{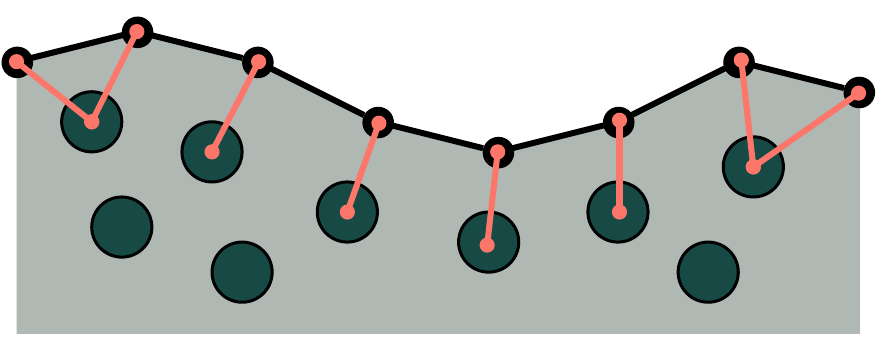}
  \caption{Illustration of the assignment of transformations to centers based on the transformations prescribed in the mesh vertices. The red lines represent the closest center for a given vertex. Transformations of vertices with the same closest center are averaged.}
  \label{fig:vertex-editing}
\end{figure}

In the case where it is more convenient to edit the sequence to deform the mesh vertices instead of the centers, it is possible first to convert the transformations prescribed to the centers to center transformations based on the closest center to the given vertex. This can be easily achieved by finding the closest center for each vertex that has a~prescribed transformation and averaging the transformations for centers with a~non-zero number of associated vertices (see Figure \ref{fig:vertex-editing}). Next, the standard method for deforming centers can be followed. However, as with center deformation, this method allows only low-frequency deformations with respect to the number of centers. Deformations that are too detailed to be described by a~set of centers cannot be performed. 

\begin{figure}[ht]
  \centering
  \includegraphics[scale=0.125]{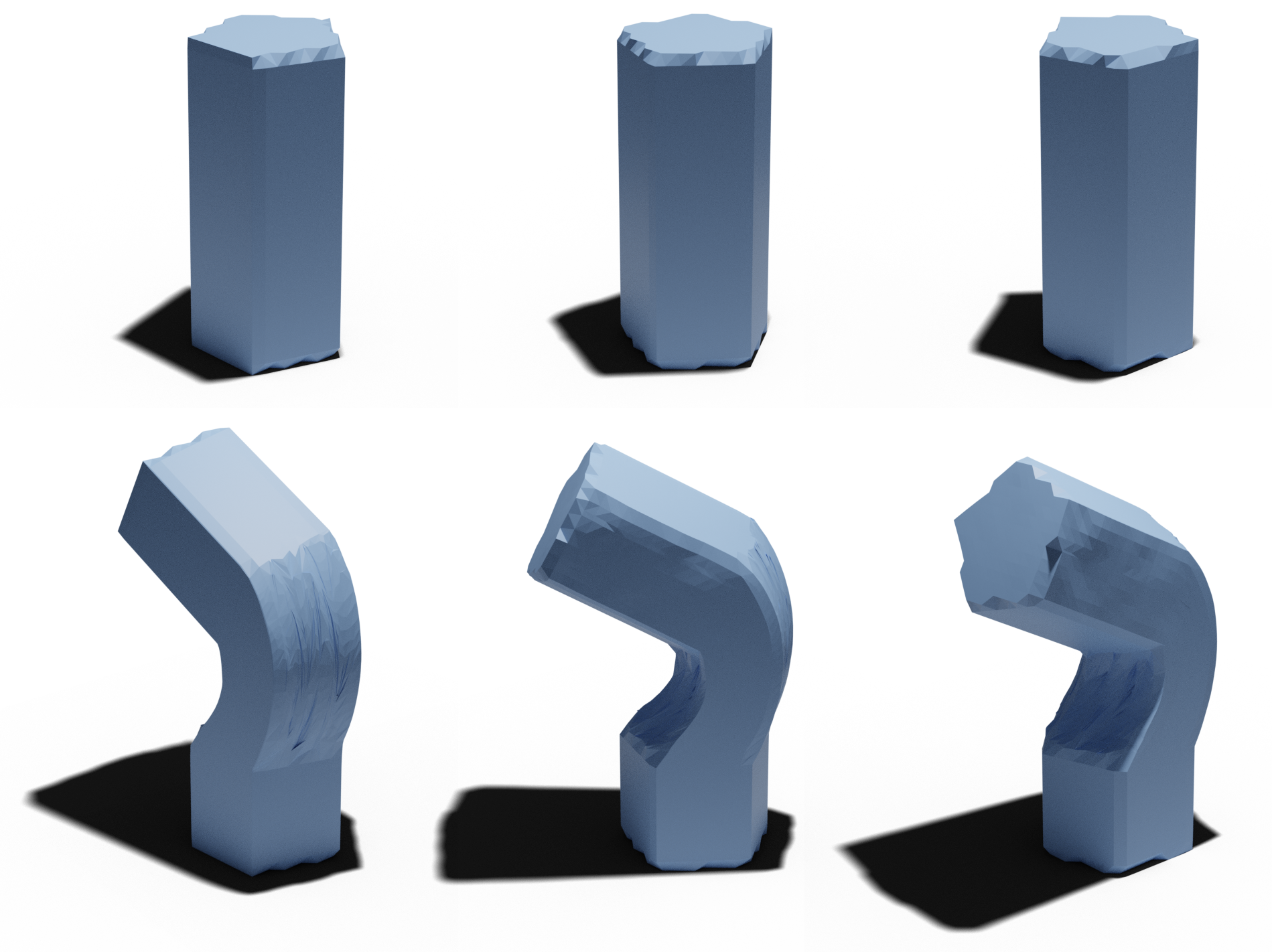}
  \caption{Example of a~sequence with rotation of a~pentahedral prism, edited by changing the position of the centers in the upper part of the shape. Top: Original sequence. Bottom: Edited sequence.}
  \label{fig:pent}
\end{figure}

\begin{figure}[ht]
  \centering
  \includegraphics[scale=0.270]{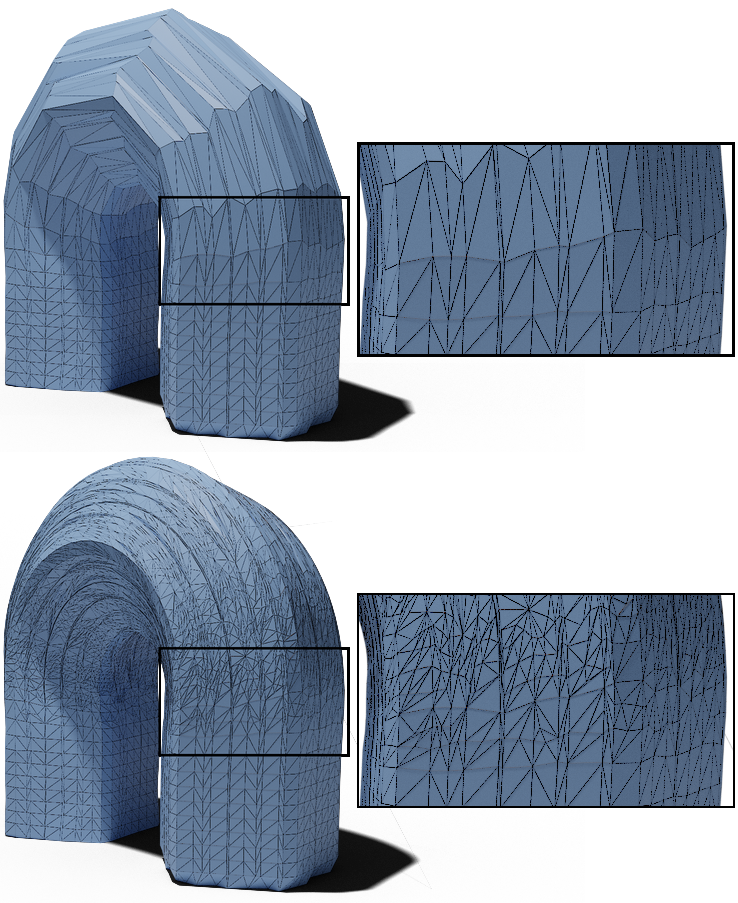}
  \caption{Comparison of deformation without and with the use of adaptive mesh subdivision. Top: Deformed without mesh subdivision. Bottom: Deformed with mesh subdivision.}
  \label{fig:subdivision}
\end{figure}

\begin{figure}[ht]
  \centering
  \includegraphics[scale=0.07]{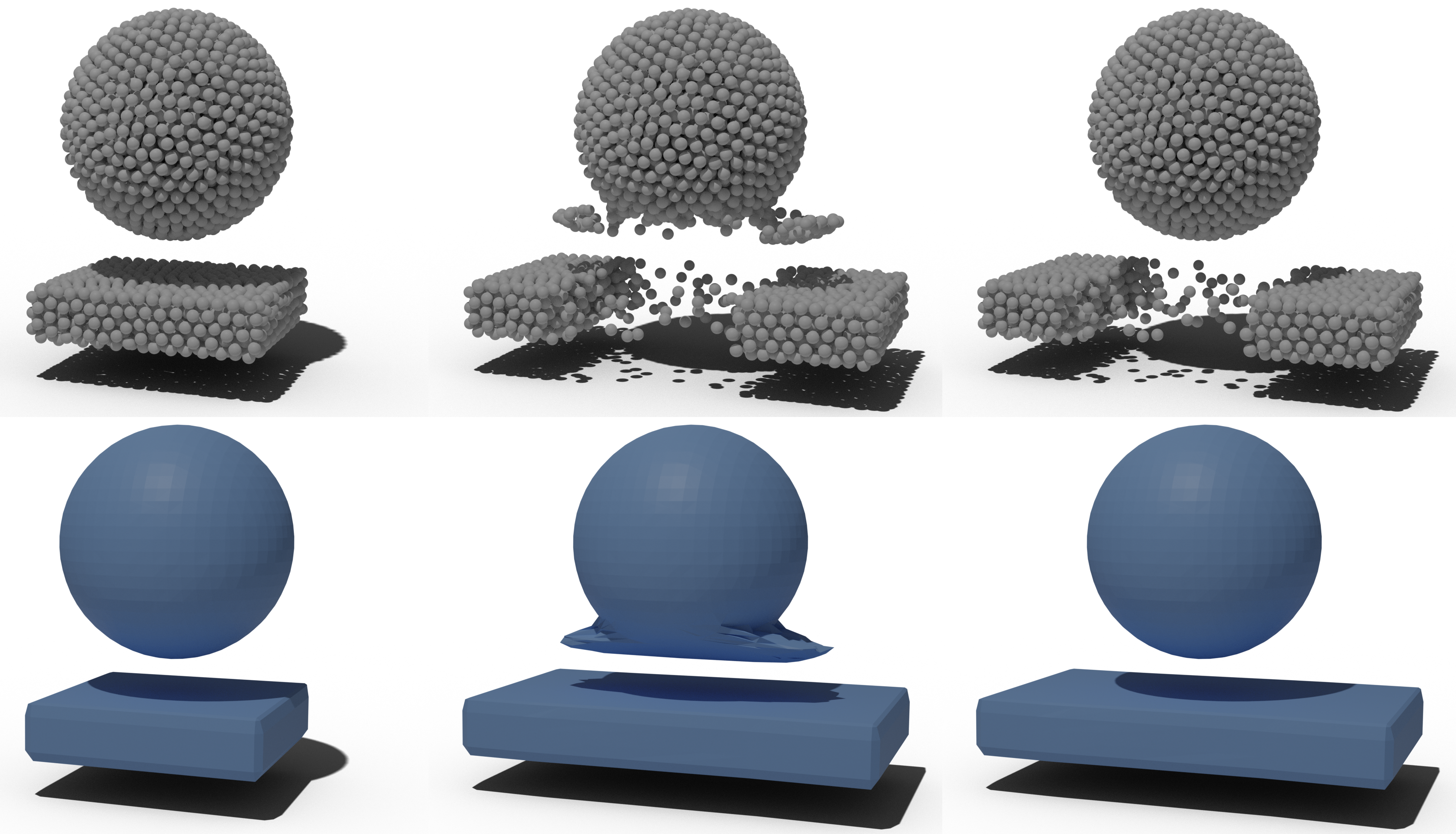}
  \caption{Comparison of editing - stretching the box component - and influence of using affinities of centers. Left: Original mesh. Middle: Edited mesh without using affinities of centers. Right: Edited mesh with the use of affinities of centers.}
  \label{fig:samba_affinities}
\end{figure}

\section{Experimental results}\label{sec:experiments}
In this section, we show the results of individual edits on several time-varying meshes. In the case of dynamic meshes (with constant connectivity), the implicit correspondences were ignored in the experiments, and thus in both cases, only the discrete correspondence given by volume tracking was used. For our experiments, we used dynamic meshes by Vlasic et al. \cite{Vlasic2008}, time-varying meshes provided by an anonymous company X, and synthetic data sets created for the purposes of demonstration. 
Note that the presented edits are intentionally strongly exaggerated in many cases, in order to demonstrate the limits of the proposed framework. Additional examples are also included in the accompanying video.

Examples of editing by locally enlarging or shrinking the volume using the center attraction are shown in Figs.~\ref{fig:att} and~\ref{fig:dis}. Fig.~\ref{fig:att} shows the volume shrinking to slim the waist on the \textit{samba} sequence. Fig.~\ref{fig:dis}, on the other hand, shows the enlarging of the volume of the head on the \textit{casual man} sequence.

Fig.~\ref{fig:pent} shows the editing by changing the position of the centers in a~rotating pentagonal prism sequence. This example shows that changing the position of the centers is consistent across the sequence, and the original character of the motion (rotation about the vertical axis) remains unchanged by the editing.

Fig.~\ref{fig:subdivision} shows a~comparison of the surface deformation with and without the use of mesh subdivision on the sequence of the pentahedral prism from Fig.~\ref{fig:pent}. The figure shows that artifacts appear in the part of the surface where the largest bending due to the specified deformation occurs. These artifacts were efficiently eliminated with the use of subdivision.

To illustrate the effect of center affinities in calculating new center positions, we show the result of editing the collision sequence (see Fig.~\ref{fig:samba_affinities}), which contains two topologically separated components. The edit consists of stretching one of the components: the box. With the propagation of deformation based on the Euclidean distance, the stretching will also cause deformation of the sphere, even though the components are separated. By using affinity based on the maximum distance of the centers in the sequence, this artifact is suppressed.

\begin{figure*}[ht]
  \centering
  \includegraphics[width=0.75\linewidth]{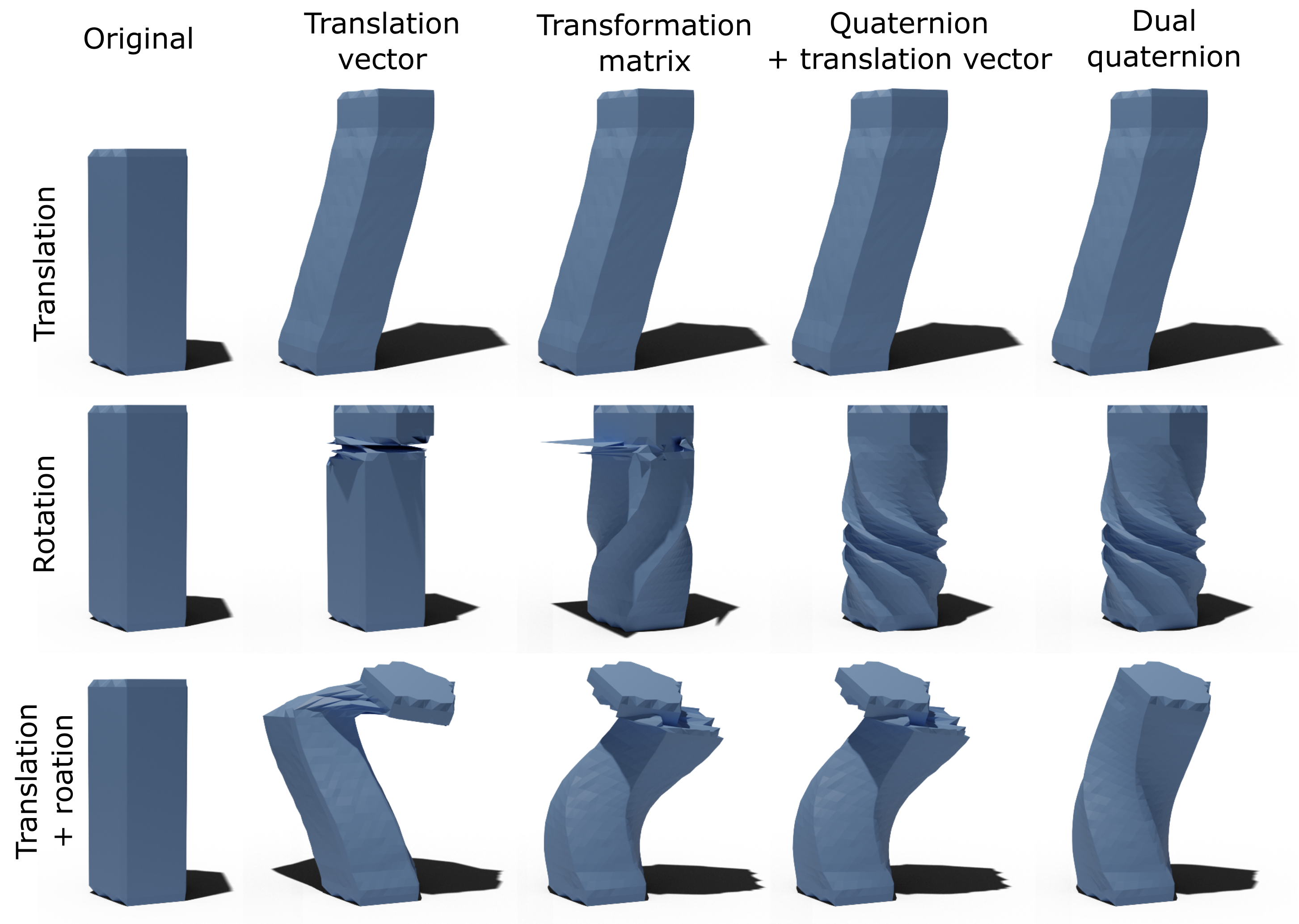}
  \caption{Comparison of different representations of rigid transformations used in the deformation model (columns) and different edits performed. Top: The top of the prism has been translated. Middle: The top of the prism has been rotated. Bottom: The top of the prism has been rotated and translated. All of these options can represent blended transforms with only a~translation component without any artifacts. For a~blend of pure rotations, the representation by real quaternions and translation vectors and the representation by dual quaternions are suitable. For the combination of translation and rotation, the best representation is the dual quaternion representation.}
  \label{fig:rotational_pent}
\end{figure*}

\begin{figure}[ht]
  \centering
  \includegraphics[scale=0.1]{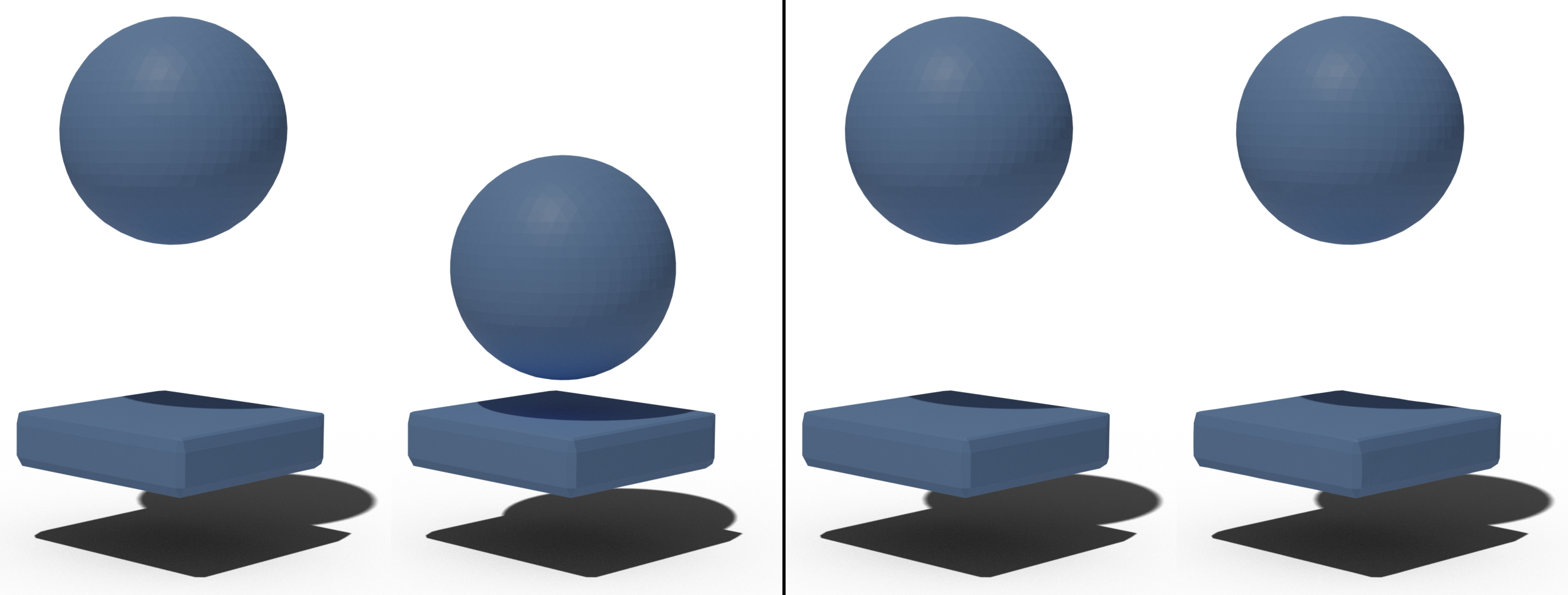}
  \caption{Example of the automatic closing of a~\textit{collision} sequence into a~loop. Left: The original pair of the first and last frames. Right: Pair of the first and the last frames after closing}
  \label{fig:ball_closing}
\end{figure}

\begin{figure}[ht]
  \centering
  \includegraphics[scale=0.125]{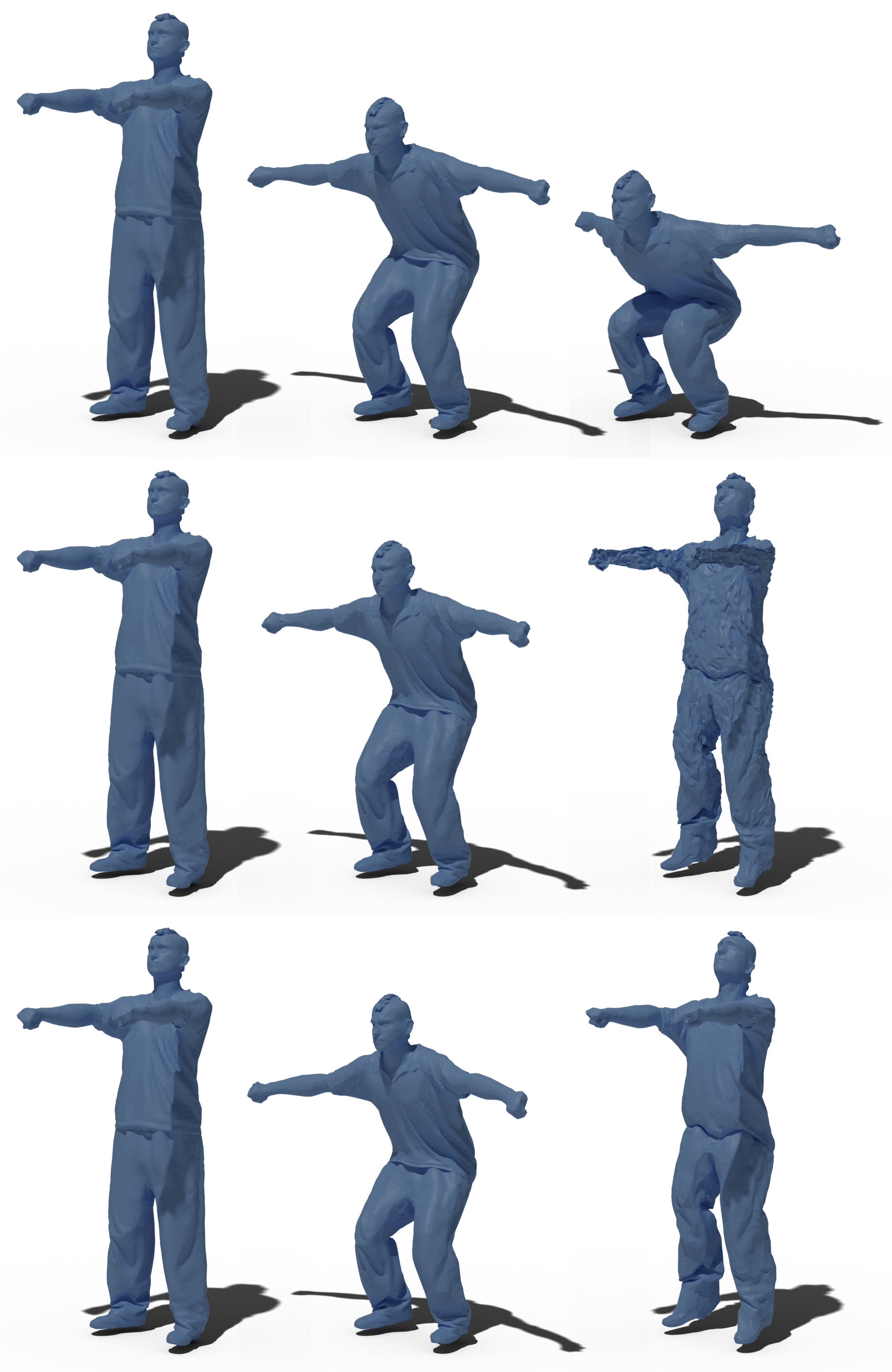}
  \caption{Example of the automatic closing of a~\textit{squat} sequence into a~loop. Top: The original meshes (the first, the middle, and the last frame). Middle: Meshes after closing using the translational model. Bottom: Meshes after closing with the dual quaternion model (including rotation components).}
  \label{fig:squat_closing}
\end{figure}

The effect of using dual quaternions as a~representation of rigid transformations compared with other representations of rigid transformations is shown in Fig.~\ref{fig:rotational_pent}. Transformations of centers can be represented using translation vectors. However, this representation completely ignores the rotational component of transformation. Another way is to represent transformations using transformation matrices. The disadvantage of this approach is that the blending of transformation matrices may not result in a~rigid transformation, and it may even be singular. It is also possible to perform the blending of rigid transformations separately for the rotational and translation parts of the transformation. The translation vectors can be blended linearly, and real quaternions can be used for blending rotations. However, the results of this approach depend on the choice of a~coordinate system. For this reason, we find dual quaternions the most suitable representation of rigid transformations. We show the results of automatically looping the sequence by comparing the pair of the first and last frames in Figs. \ref{fig:ball_closing} and \ref{fig:squat_closing}. The remaining surface distortion seen in Figure \ref{fig:squat_closing} is due to the inaccuracy of the volume tracking.
To minimize these artifacts, it is crucial that the distances of the centers from their affine surroundings change as little as possible. Otherwise, the error is accumulated, which is most noticeable in the case of sequence looping.

\begin{figure*}[ht]
  \centering
  \includegraphics[scale=0.175]{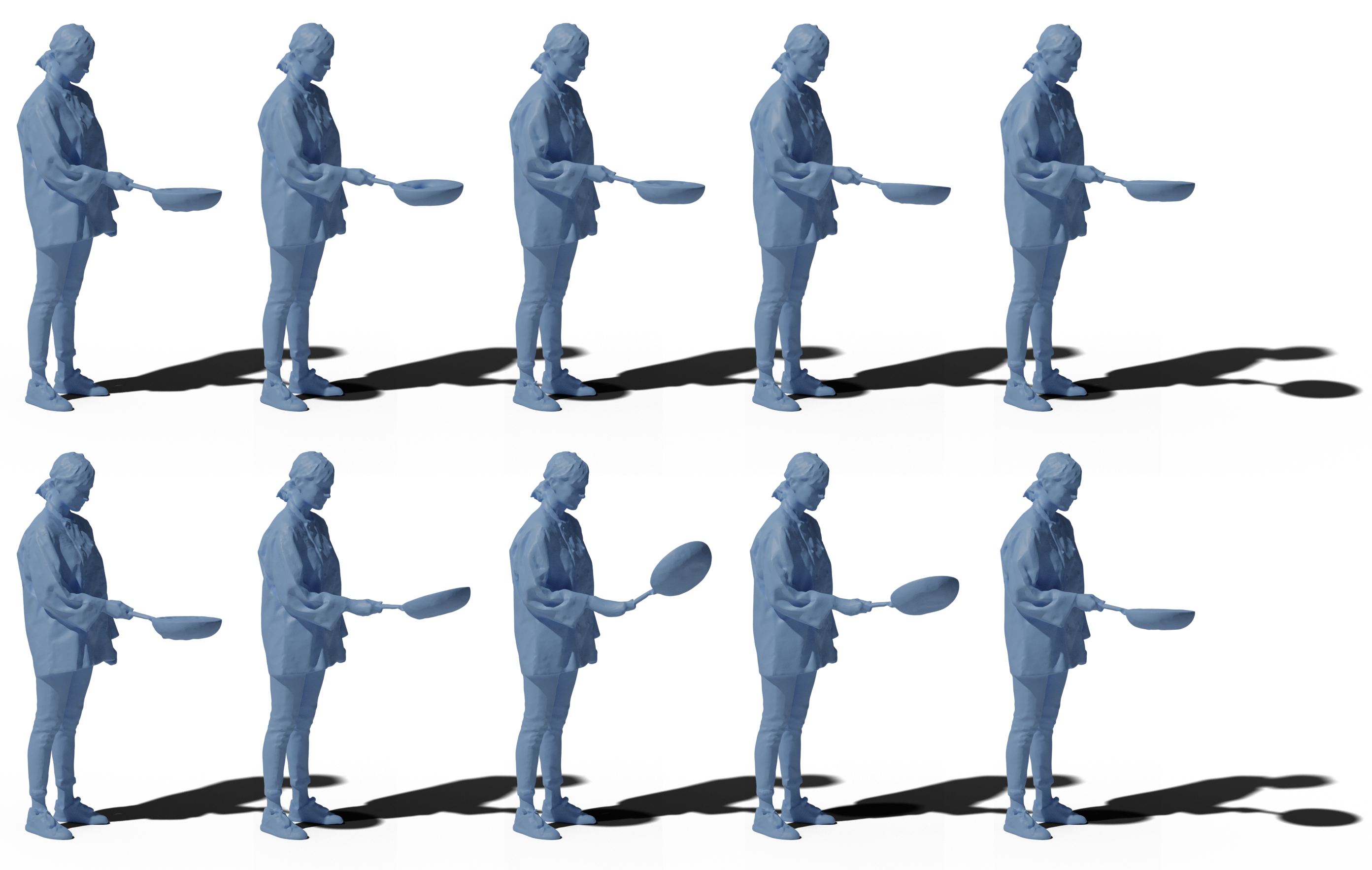}
  \caption{Example of a~sequence edited by changing the position of the centers - tilting up the pan. Top: Original sequence. Bottom:Edited sequence.}
  \label{fig:cooking}
\end{figure*}

Fig.~\ref{fig:cooking} shows the possibility of creating a~new movement on a~\textit{cooking} sequence. The original sub-sequence shows an almost static surface. The new sequence was created by changing the position of the centers inside the pan and then propagating this edit using \textit{time decay}. In the electronic version of the paper, the sequence is shown in the supplementary video file.

\begin{figure}[ht]
  \centering
  \includegraphics[scale=0.105]{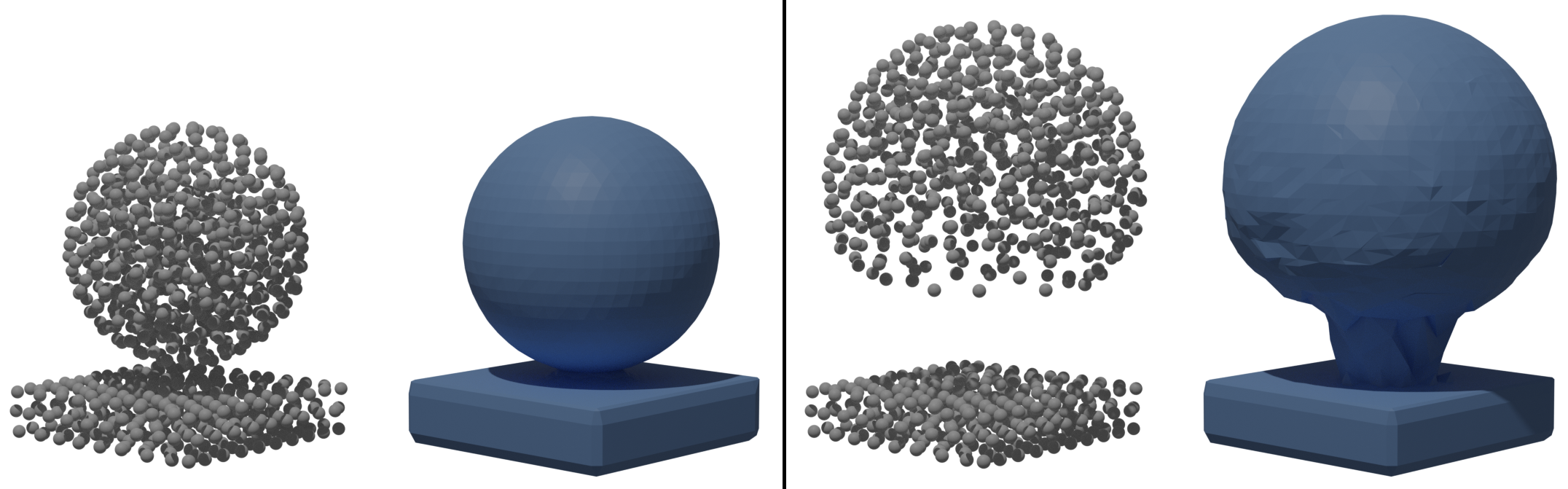}
  \caption{Example of problematic edits near a~hidden surface. It is not possible to separate the components without changing the mesh connectivity. Left: Original mesh frame. Right: Edited mesh frame.}
  \label{fig:topology}
\end{figure}

A problematic part of TVM editing is editing parts of the surface near topological noise. Fig.~\ref{fig:topology} shows an attempt to separate two components in a~collision sequence, in which a~part of the surface was hidden between touching components. In terms of volume tracking and computed center affinities, it is obvious that these are two different components. Since the surface topology is not correct in this frame of the sequence, and the editing framework does not allow for a change in topology, since it preserves the original mesh connectivity up to a subdivision, the components cannot be correctly separated.

\begin{figure}[ht]
  \centering
  \includegraphics[width=0.99\linewidth]{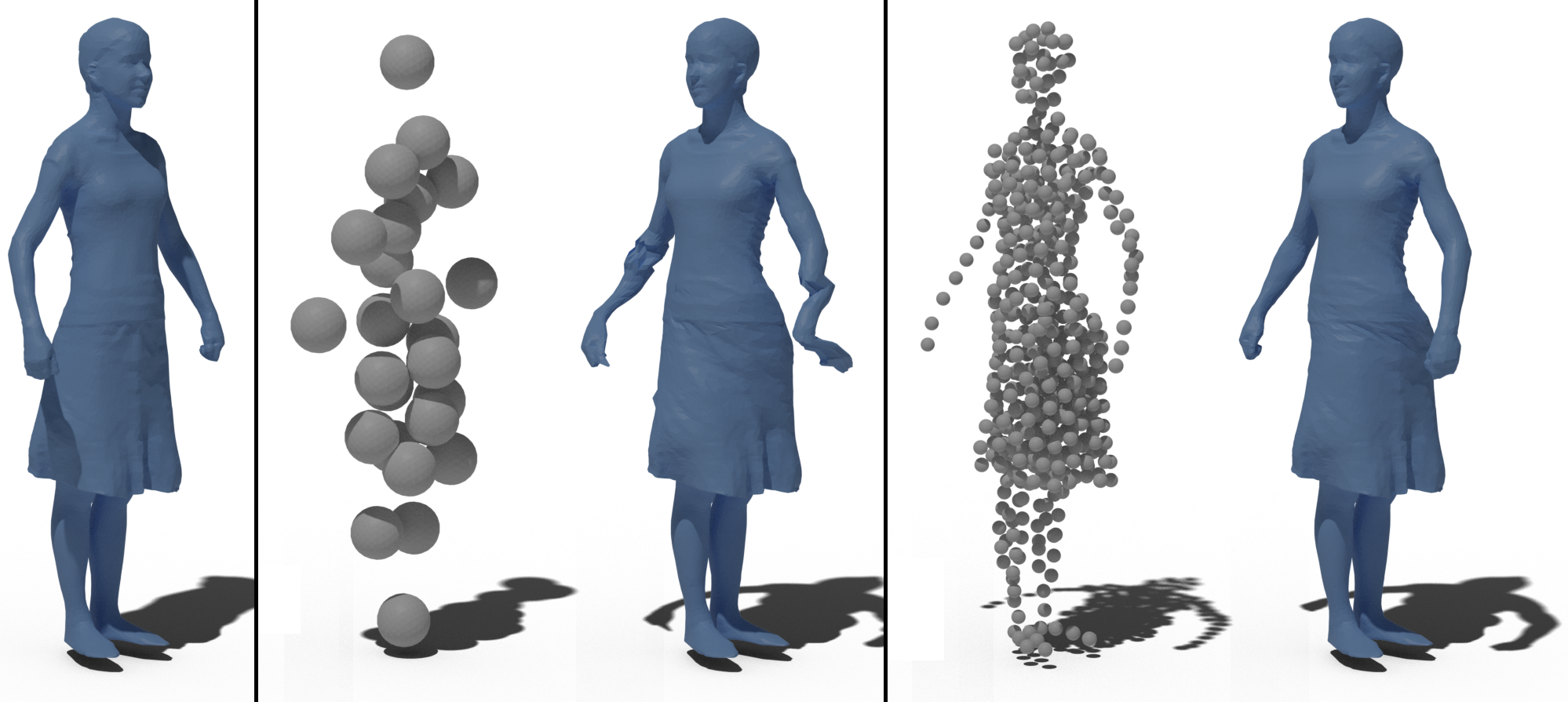}
  \caption{Ablation study of the number of tracked centers. An insufficient number of centers to represent the edited shape leads to undesirable artifacts. Left: Original shape. Middle: Quarter turn edit using 25 centers. Right: Quarter turn edit using 500 centers.}
  \label{fig:centers_num}
\end{figure}

An important parameter for editing sequences is the density of tracked centers. It is essential that the number of centers is sufficient to represent the low-frequency geometry of the deformed shape and its trajectory (see Fig.~\ref{fig:centers_num}). However, the number of centers used significantly affects the tracking time since the dependence of the tracking time on the number of centers is quadratic. For this reason, choosing a~trade-off between computational complexity and shape expressiveness of the centers is necessary. The number of centers used in these experiments ranged between 1000 and 4000.

The editing time is dependent on the number of mesh vertices edited, the number of frames in the sequence, and the number of tracked centers. The most time-consuming part is the computation of the deformation weights of the vertices. However, this step is performed only once. For subsequent edits, the precomputed weights can be used. We measured the computation time on a~configuration with Intel Core i9-11900K 3.5 GHz, 64~GB RAM, with all computations performed on the CPU. The time required to compute the center deformations in the single frame edited and the following propagation into the centers of the remaining frames on the~\textit{casual man} sequence, spanning 545 frames and averaging 36,147 vertices per frame, with 4,000 volume centers tracked, was 25,724~ms (47~ms per frame). The time required for the surface deformation was 124,283~ms (228 ms per frame). Once the surface weights had been calculated, the next surface editing was sped up to 7273~ms (13~ms per frame), which resulted in a~total cost of 49.1 ms per frame with precomputed weights for surface editing, thus allowing for real-time visualization of the results.

When calculating affinities, it is necessary to determine the maximum distance to all other centers for each center. The complexity of this calculation is $\mathcal{O}(N^2F)$, where $N$ denotes the number of centers and $F$ denotes the number of frames in the sequence. With an efficient data structure for nearest point evaluation the complexity is $\Theta(NlogNF)$ for practical cases. When propagating the editing of $E$ centers in a single frame of the sequence to the other centers, a weighted sum for each deformed center needs to be computed for each of the other centers. The complexity of this step is therefore $\Theta(EN)\subset \mathcal{O}(N^2)$. Propagating the deformation to the $F-1$ other frames of the sequence requires finding the $k$-nearest neighbors for each center with complexity $\Theta(N^2)$, but this needs to be performed only once. Assuming that the number of $k$ neighbors is significantly smaller than the total number of centers $N$, the complexity of propagation to the other frames of the sequence is $\Theta(NF)$. 
When deforming the surface of the sequence itself with an average number V of vertices per frame without mesh resampling, for each vertex, its nearest center and the next $k-1$ most affine neighbors need to be found to determine the weights. This operation also needs to be performed only once. The complexity of mesh deformation itself is $\Theta(VF)$. When using adaptive mesh subdivision with a maximum number of iterations $S$, the number of new vertices is limited from above by the number of edges, which is, according to Euler's formula, approximately three times more than the number vertices in a triangle mesh. Therefore, the complexity of deforming meshes with subdivision is bounded by the complexity of $\mathcal{O}(4^SVF)$.
\section{Conclusions}
We present a~time-consistent editing framework for time-varying meshes using discrete (sparse) correspondences obtained by a readily available volume tracking method. The editing framework solves three main problems: editing centers in one frame, propagating the editing to other frames, and deforming the surface triangle meshes by the centers. The framework supports three operations: rigid transformation of selected centers, inflation/deflation, and looping the sequence. We have proposed a~method to compute weights for editing triangle meshes satisfying the required conditions and an adaptive mesh subdivision mechanism to preserve the sampling density of the surface. The proposed methods have been tested and evaluated on various time-varying meshes. We have shown the benefits of using dual quaternion blending for deformation interpolation and the importance of incorporating the rotation component of the transformations within the deformation model. 

The proposed method is applicable in scenarios where minor changes are required, such as modifying the proportions of a character. It provides mostly plausible results, fully automatically, even up to the exaggerated cases, and thus saves considerable effort otherwise necessary when each frame must be edited separately. 

Naturally, the quality of the result is strongly related to the success of the volume tracking step. Any inaccuracy in the positions of the tracked centers causes undesired distortion of the surface during subsequent TVM editing. In particular, the most significant problems are encountered when a~tracked center incorrectly skips from one connected component of the tracked surface to another one due to self-contact. We are currently using the publicly available implementation of volume center tracking, and any future improvement to it is going to improve the editing method as well.

An essential component of the deformation model is the affinity of centers, which allows distinguishing between Euclidean and geodesically close centers. This, in turn, allows the editing result to reflect the information from the whole sequence. If some centers become separated at any point, the concept of affinity is going to capture this information and use it to propagate the edits naturally. However, this distinction is possible only when these centers are sufficiently distant from each other during the sequence. Future work could further address this problem. Separation of centers that are not distant from each other during the whole sequence, although they are topologically separate, would be possible, e.g., using additional input from the user by specifying a precise description of the topology or setting fixed affinities for key pairs of centers.

Currently, the method is limited to edits on the scale defined by the tracked centers, and therefore, it does not allow editing fine details. Our deformation model could potentially be used to derive a surface correspondence, which in turn may allow editing on a smaller scale. Additionally, the visual quality of the deformations strongly depends on the quality of the estimated correspondences using volume tracking or other methods.

Another limitation is the ability to edit body parts near a~hidden surface. A~consequence of the hidden surface is the incorrect topology of such mesh frames. While maintaining connectivity, such parts cannot be edited correctly. In future work, it would be useful to extend the framework to allow local editing of connectivity in such a~way that the resulting connectivity matches the topology estimated from the affinities of the tracked centers.
The source code of our framework, along with videos of edited sequences, is publicly available at \url{https://gitlab.com/hachaf/tvm-editing}.


\bibliographystyle{abbrv}
\bibliography{refs}

\begin{thebibliography}{10}

\bibitem{Adams2008}
B.~Adams, M.~Ovsjanikov, M.~Wand, H.-P. Seidel, and L.~J. Guibas.
\newblock Meshless modeling of deformable shapes and their motion.
\newblock In {\em Proceedings of the 2008 ACM SIGGRAPH/Eurographics Symposium on Computer Animation}, SCA '08, page 77–86, Goslar, DEU, 2008. Eurographics Association.

\bibitem{barbic}
J.~Barbi\v{c}, F.~Sin, and E.~Grinspun.
\newblock Interactive editing of deformable simulations.
\newblock {\em ACM Trans. Graph.}, 31(4), jul 2012.

\bibitem{bojsen2012tracking}
M.~Bojsen-Hansen, H.~Li, and C.~Wojtan.
\newblock Tracking surfaces with evolving topology.
\newblock {\em ACM Transactions on Graphics (Proceedings SIGGRAPH 2012)}, 31(4), 2012.

\bibitem{Boltzmann1868}
L.~Boltzmann.
\newblock {\em Studien {\"u}ber das Gleichgewicht der lebendigen Kraft zwischen bewegten materiellen Punkten: vorgelegt in der Sitzung am 8. October 1868}.
\newblock k. und k. Hof- und Staatsdr., 1868.

\bibitem{pmp}
M.~Botsch, L.~Kobbelt, M.~Pauly, P.~Alliez, and B.~L{\'e}vy.
\newblock {\em Polygon mesh processing}.
\newblock CRC press, 2010.

\bibitem{primo}
M.~Botsch, M.~Pauly, M.~Gross, and L.~Kobbelt.
\newblock {PriMo: Coupled Prisms for Intuitive Surface Modeling}.
\newblock In A.~Sheffer and K.~Polthier, editors, {\em Symposium on Geometry Processing}. The Eurographics Association, 2006.

\bibitem{variational}
M.~Botsch and O.~Sorkine.
\newblock On linear variational surface deformation methods.
\newblock {\em IEEE Transactions on Visualization and Computer Graphics}, 14(1):213--230, 2008.

\bibitem{casas1}
D.~Casas, M.~Tejera, J.-Y. Guillemaut, and A.~Hilton.
\newblock Parametric control of captured mesh sequences for real-time animation.
\newblock In J.~M. Allbeck and P.~Faloutsos, editors, {\em Motion in Games}, pages 242--253, Berlin, Heidelberg, 2011. Springer Berlin Heidelberg.

\bibitem{casas2}
D.~Casas, M.~Tejera, J.-Y. Guillemaut, and A.~Hilton.
\newblock 4d parametric motion graphs for interactive animation.
\newblock In {\em Proceedings of the ACM SIGGRAPH Symposium on Interactive 3D Graphics and Games}, I3D '12, page 103–110, New York, NY, USA, 2012. Association for Computing Machinery.

\bibitem{casas3}
D.~Casas, M.~Tejera, J.-Y. Guillemaut, and A.~Hilton.
\newblock Interactive animation of 4d performance capture.
\newblock {\em IEEE Transactions on Visualization and Computer Graphics}, 19(5):762--773, 2013.

\bibitem{cashman}
T.~J. Cashman and K.~Hormann.
\newblock A continuous, editable representation for deforming mesh sequences with separate signals for time, pose and shape.
\newblock {\em Computer Graphics Forum}, 31(2pt4):735--744, 2012.

\bibitem{aguiarskinning}
E.~De~Aguiar, C.~Theobalt, S.~Thrun, and H.-P. Seidel.
\newblock Automatic conversion of mesh animations into skeleton-based animations.
\newblock {\em Computer Graphics Forum}, 27(2):389--397, 2008.

\bibitem{aguiar}
E.~de~Aguiar and N.~Ukita.
\newblock Representing and manipulating mesh-based character animations.
\newblock In {\em 2012 25th SIBGRAPI Conference on Graphics, Patterns and Images}, pages 198--204, 2012.

\bibitem{magic_wand}
C.~J. Dean, J.~P. Lewis, and A.~Chalmers.
\newblock A magic wand for motion capture editing and edit propagation.
\newblock In {\em SIGGRAPH Asia 2018 Technical Briefs}, SA '18, New York, NY, USA, 2018. Association for Computing Machinery.

\bibitem{doumanoglou}
A.~Doumanoglou, D.~S. Alexiadis, D.~Zarpalas, and P.~Daras.
\newblock Toward real-time and efficient compression of human time-varying meshes.
\newblock {\em IEEE Transactions on Circuits and Systems for Video Technology}, 24(12):2099--2116, 2014.

\bibitem{Dvorak2023}
J.~Dvo{\v{r}}{\'a}k, F.~H{\'a}cha, and L.~V{\'a}{\v{s}}a.
\newblock Global optimisation for improved volume tracking of time-varying meshes.
\newblock In J.~Miky{\v{s}}ka, C.~de~Mulatier, M.~Paszynski, V.~V. Krzhizhanovskaya, J.~J. Dongarra, and P.~M. Sloot, editors, {\em Computational Science -- ICCS 2023}, pages 113--127, Cham, 2023. Springer Nature Switzerland.

\bibitem{Dvorak2021}
J.~Dvo\v{r}\'{a}k, P.~Van\v{e}\v{c}ek, and L.~V\'{a}\v{s}a.
\newblock Towards understanding time varying triangle meshes.
\newblock In {\em Computational Science – ICCS 2021: 21st International Conference, Krakow, Poland, June 16–18, 2021, Proceedings, Part V}, page 45–58, Berlin, Heidelberg, 2021. Springer-Verlag.

\bibitem{Dvorak2022}
J.~Dvořák, Z.~Káčereková, P.~Vaněček, L.~Hruda, and L.~Váša.
\newblock As-rigid-as-possible volume tracking for time-varying surfaces.
\newblock {\em Computers \& Graphics}, 102:329--338, 2022.

\bibitem{divfree}
M.~Eisenberger, Z.~Lähner, and D.~Cremers.
\newblock Divergence‐free shape correspondence by deformation.
\newblock {\em Computer Graphics Forum}, 38:1--12, 08 2019.

\bibitem{abrevaya}
V.~{Fernández Abrevaya}, S.~Manandhar, F.~Hétroy-Wheeler, and S.~Wuhrer.
\newblock A 3d+t laplace operator for temporal mesh sequences.
\newblock {\em Computers \& Graphics}, 58:12--22, 2016.
\newblock Shape Modeling International 2016.

\bibitem{gleicher97}
M.~Gleicher.
\newblock Motion editing with spacetime constraints.
\newblock In {\em Proceedings of the 1997 Symposium on Interactive 3D Graphics}, I3D '97, page 139–ff., New York, NY, USA, 1997. Association for Computing Machinery.

\bibitem{heck}
R.~Heck and M.~Gleicher.
\newblock Parametric motion graphs.
\newblock In {\em Proceedings of the 2007 Symposium on Interactive 3D Graphics and Games}, I3D '07, page 129–136, New York, NY, USA, 2007. Association for Computing Machinery.

\bibitem{hildebrandt}
K.~Hildebrandt, C.~Schulz, C.~von Tycowicz, and K.~Polthier.
\newblock Interactive spacetime control of deformable objects.
\newblock {\em ACM Trans. Graph.}, 31(4), 2012.

\bibitem{huang_tracking}
C.-H.~P. Huang, B.~Allain, E.~Boyer, J.-S. Franco, F.~Tombari, N.~Navab, and S.~Ilic.
\newblock Tracking-by-detection of 3d human shapes: From surfaces to volumes.
\newblock {\em IEEE Transactions on Pattern Analysis and Machine Intelligence}, 40(8):1994--2008, 2018.

\bibitem{huang_graphs}
P.~Huang, M.~Tejera, J.~Collomosse, and A.~Hilton.
\newblock Hybrid skeletal-surface motion graphs for character animation from 4d performance capture.
\newblock {\em ACM Trans. Graph.}, 34(2), mar 2015.

\bibitem{biharmonic}
A.~Jacobson, I.~Baran, J.~Popovi\'{c}, and O.~Sorkine.
\newblock Bounded biharmonic weights for real-time deformation.
\newblock {\em ACM Trans. Graph.}, 30(4), jul 2011.

\bibitem{dmskinning}
D.~L. James and C.~D. Twigg.
\newblock Skinning mesh animations.
\newblock {\em ACM Trans. Graph.}, 24(3):399–407, jul 2005.

\bibitem{animesh}
M.~Jin, D.~Gopstein, Y.~Gingold, and A.~Nealen.
\newblock Animesh: Interleaved animation, modeling, and editing.
\newblock {\em ACM Trans. Graph.}, 34(6), nov 2015.

\bibitem{Kabsch1976}
W.~Kabsch.
\newblock {A solution for the best rotation to relate two sets of vectors}.
\newblock {\em Acta Crystallographica Section A}, 32(5):922--923, 1976.

\bibitem{Kavan2008}
L.~Kavan, S.~Collins, J.~\v{Z}\'{a}ra, and C.~O'Sullivan.
\newblock Geometric skinning with approximate dual quaternion blending.
\newblock {\em ACM Trans. Graph.}, 27(4), nov 2008.

\bibitem{kim2007motion}
H.~B. Kim and H.~Kim.
\newblock Motion editing based on joint classification and discrete blending.
\newblock {\em IJCSNS}, 7(2):55, 2007.

\bibitem{kircher_editing}
S.~Kircher and M.~Garland.
\newblock Editing arbitrarily deforming surface animations.
\newblock {\em ACM Trans. Graph.}, 25(3):1098–1107, jul 2006.

\bibitem{kircher_free_form}
S.~Kircher and M.~Garland.
\newblock Free-form motion processing.
\newblock {\em ACM Trans. Graph.}, 27(2), may 2008.

\bibitem{kovar}
L.~Kovar, M.~Gleicher, and F.~Pighin.
\newblock Motion graphs.
\newblock In {\em ACM SIGGRAPH 2008 Classes}, SIGGRAPH '08, New York, NY, USA, 2008. Association for Computing Machinery.

\bibitem{optimo}
Y.~Koyama and M.~Goto.
\newblock Optimo: Optimization-guided motion editing for keyframe character animation.
\newblock In {\em Proceedings of the 2018 CHI Conference on Human Factors in Computing Systems}, CHI '18, page 1–12, New York, NY, USA, 2018. Association for Computing Machinery.

\bibitem{rigging}
B.~H. Le and Z.~Deng.
\newblock Robust and accurate skeletal rigging from mesh sequences.
\newblock {\em ACM Trans. Graph.}, 33(4), jul 2014.

\bibitem{skeleton_laplacian}
T.~Le~Naour, N.~Courty, and S.~Gibet.
\newblock Spatiotemporal coupling with the 3d+t motion laplacian.
\newblock {\em Computer Animation and Virtual Worlds}, 24(3-4):419--428, 2013.

\bibitem{lisimulation2}
S.~Li, J.~Huang, F.~de~Goes, X.~Jin, H.~Bao, and M.~Desbrun.
\newblock Space-time editing of elastic motion through material optimization and reduction.
\newblock {\em ACM Trans. Graph.}, 33(4), jul 2014.

\bibitem{lisimulation1}
S.~Li, J.~Huang, M.~Desbrun, and X.~Jin.
\newblock Interactive elastic motion editing through space–time position constraints.
\newblock {\em Computer Animation and Virtual Worlds}, 24(3-4):409--417, 2013.

\bibitem{liysimulation}
Y.~Li, H.~Xu, and J.~Barbič.
\newblock Enriching triangle mesh animations with physically based simulation.
\newblock {\em IEEE Transactions on Visualization and Computer Graphics}, 23(10):2301--2313, 2017.

\bibitem{prada}
F.~Prada, M.~Kazhdan, M.~Chuang, A.~Collet, and H.~Hoppe.
\newblock Motion graphs for unstructured textured meshes.
\newblock {\em ACM Trans. Graph.}, 35(4), jul 2016.

\bibitem{killingfusion}
M.~Slavcheva, M.~Baust, D.~Cremers, and S.~Ilic.
\newblock Killingfusion: Non-rigid 3d reconstruction without correspondences.
\newblock In {\em 2017 IEEE Conference on Computer Vision and Pattern Recognition (CVPR)}, pages 5474--5483, 2017.

\bibitem{arap}
O.~Sorkine and M.~Alexa.
\newblock As-rigid-as-possible surface modeling.
\newblock In {\em Proceedings of the Fifth Eurographics Symposium on Geometry Processing}, SGP '07, page 109–116, Goslar, DEU, 2007. Eurographics Association.

\bibitem{lapedit}
O.~Sorkine, D.~Cohen-Or, Y.~Lipman, M.~Alexa, C.~R\"{o}ssl, and H.-P. Seidel.
\newblock Laplacian surface editing.
\newblock In {\em Proceedings of the 2004 Eurographics/ACM SIGGRAPH Symposium on Geometry Processing}, SGP '04, page 175–184, New York, NY, USA, 2004. Association for Computing Machinery.

\bibitem{SorkineRabinovich:SVD-rotations:2016}
O.~Sorkine-Hornung and M.~Rabinovich.
\newblock Least-squares rigid motion using svd, 2016.
\newblock Technical note.

\bibitem{Sumner2007}
R.~W. Sumner, J.~Schmid, and M.~Pauly.
\newblock Embedded deformation for shape manipulation.
\newblock {\em ACM Trans. Graph.}, 26(3):80–es, jul 2007.

\bibitem{tejera2}
M.~Tejera, D.~Casas, and A.~Hilton.
\newblock Animation control of surface motion capture.
\newblock {\em IEEE Transactions on Cybernetics}, 43(6):1532--1545, 2013.

\bibitem{tejera1}
M.~Tejera and A.~Hilton.
\newblock Space-time editing of 3d video sequences.
\newblock In {\em 2011 Conference for Visual Media Production}, pages 148--157, 2011.

\bibitem{tevs}
A.~Tevs, A.~Berner, M.~Wand, I.~Ihrke, M.~Bokeloh, J.~Kerber, and H.-P. Seidel.
\newblock Animation cartography—intrinsic reconstruction of shape and motion.
\newblock {\em ACM Trans. Graph.}, 31(2), apr 2012.

\bibitem{cager}
J.-M. Thiery, J.~Tierny, and T.~Boubekeur.
\newblock Cager: Cage-based reverse engineering of animated 3d shapes.
\newblock {\em Computer Graphics Forum}, 31(8):2303--2316, 2012.

\bibitem{Vlasic2008}
D.~Vlasic, I.~Baran, W.~Matusik, and J.~Popovi\'{c}.
\newblock Articulated mesh animation from multi-view silhouettes.
\newblock {\em ACM Trans. Graph.}, 27(3):1–9, aug 2008.

\bibitem{STED}
L.~Váša and V.~Skala.
\newblock A perception correlated comparison method for dynamic meshes.
\newblock {\em IEEE transactions on visualization and computer graphics}, 17:220--30, 02 2011.

\bibitem{wu2008space}
Y.~Wu, H.~Zhang, C.~Song, and H.~Bao.
\newblock Space-time curve analogies for motion editing.
\newblock In {\em International Conference on Geometric Modeling and Processing}, pages 437--449. Springer, 2008.

\bibitem{Xu2008}
J.~Xu, T.~Yamasaki, and K.~Aizawa.
\newblock Motion editing for time-varying mesh.
\newblock {\em EURASIP Journal on Advances in Signal Processing}, 2009, 03 2008.

\bibitem{xu_gradient}
W.~Xu, K.~Zhou, Y.~Yu, Q.~Tan, Q.~Peng, and B.~Guo.
\newblock Gradient domain editing of deforming mesh sequences.
\newblock {\em ACM Trans. Graph.}, 26(3):84–es, jul 2007.

\bibitem{tvmeditingyang}
L.~Yang, C.~Xiao, and J.~Fang.
\newblock Multi-scale geometric detail enhancement for time-varying surfaces.
\newblock {\em Graphical Models}, 76(5):413--425, 2014.
\newblock Geometric Modeling and Processing 2014.

\bibitem{neural_cages}
W.~Yifan, N.~Aigerman, V.~G. Kim, S.~Chaudhuri, and O.~Sorkine-Hornung.
\newblock Neural cages for detail-preserving 3d deformations.
\newblock In {\em 2020 IEEE/CVF Conference on Computer Vision and Pattern Recognition (CVPR)}, pages 72--80, 2020.

\bibitem{edit_survey}
Y.-J. Yuan, Y.-K. Lai, T.~Wu, L.~Gao, and L.~Liu.
\newblock A revisit of shape editing techniques: From the geometric to the neural viewpoint.
\newblock {\em Journal of Computer Science and Technology}, 36(3):520--554, 2021.

\bibitem{zhao}
Y.~Zhao, H.~Qian, and S.~Lu.
\newblock A deformation-aware hierarchical framework for shape-preserving editing of static and time-varying mesh data.
\newblock {\em Computers \& Graphics}, 46:80--88, 2015.
\newblock Shape Modeling International 2014.

\end{thebibliography}
\end{document}